\documentclass[pra,aps,preprint,showpacs,superscriptaddress]{revtex4-1}
\usepackage{mathrsfs}
\usepackage{amsmath}
\usepackage{amssymb}
\usepackage{graphicx}
\usepackage{esint}
\begin{document}

\def\be{\begin{equation}}
\def\ee{\end{equation}}

\title{Bias-modulated dynamics of a strongly driven two-level system}

\author{Zhiguo L\"{u}}
\email{zglv@sjtu.edu.cn}
\affiliation{Key Laboratory of Artificial Structures and Quantum Control (Ministry of Education), Department of Physics and Astronomy, Shanghai Jiao Tong University, Shanghai, 200240, China}
\affiliation{Innovation Center of Advanced Microstructures, Nanjing
  210093, China}
\author{Yiying Yan}
\affiliation{Key Laboratory of Artificial Structures and Quantum Control (Ministry of Education), Department of Physics and Astronomy, Shanghai Jiao Tong University, Shanghai, 200240, China}
\affiliation{Innovation Center of Advanced Microstructures, Nanjing
  210093, China}
\author{Hsi-Sheng Goan}
\email{goan@phys.ntu.edu.tw}
\affiliation{Department of Physics and Center for Theoretical
  Sciences, National Taiwan University, Taipei 10617 }
\affiliation{Center for Quantum Science and Engineering, National Taiwan University, Taipei 10617, Taiwan}
\author{Hang Zheng}
\email{hzheng@sjtu.edu.cn}
\affiliation{Key Laboratory of Artificial Structures and Quantum Control (Ministry of Education), Department of Physics and Astronomy, Shanghai Jiao Tong University, Shanghai, 200240, China}
\affiliation{Innovation Center of Advanced Microstructures, Nanjing
  210093, China}

\date{\today}
\begin{abstract}
We investigate the bias-modulated dynamics of a strongly driven
two-level system using the counter-rotating-hybridized rotating-wave
(CHRW) method. This CHRW method treats the driving field and the bias on equal footing by a unitary transformation
with two parameters $\xi$ and $\zeta$, and is nonperturbative in
driving strength, tunneling amplitude or bias. In addition, this CHRW method is
beyond the traditional rotating-wave approximation (Rabi-RWA) and yet
by properly choosing the two parameters $\xi$ and $\zeta$,
the transformed Hamiltonian takes the RWA form with a
renormalized energy splitting and a renormalized driving
strength. The reformulated CHRW  method possesses the same
mathematical simplicity as the Rabi-RWA approach and thus allows us to
calculate analytically the dynamics and explore explicitly the effect
of the bias.
We show that the CHRW method gives the
accurate driven dynamics for a wide
range of parameters as compared to the numerically exact results.
When energy scales of the driving are comparable to
the intrinsic energy scale of the two-level systems, the counter-rotating
interactions and static bias profoundly influence the generalized Rabi
frequency. In this regime, where ordinary perturbation approaches fail,
the CHRW works very well and efficiently.
We also demonstrate the dynamics of the system in the
strong-driving and off-resonance cases for which the Rabi-RWA method
breaks down but the CHRW method remains valid.
We obtain analytical expressions for the
generalized Rabi frequency and bias-modulated Bloch-Siegert shift as
functions of the bias, tunneling and driving field
parameters. The CHRW approach is a mathematically simple and physically clear method.
It can be applied to treat some complicated
problems for which a numerical study is difficult to perform.

\end{abstract}


\maketitle





\section{INTRODUCTION} \label{intro}
Quantum mechanically two-level systems (TLSs) provide an ideal testing
ground for exploring nonclassical phenomena and understanding the
nature of quantum physics \cite{rmp,book}. The primary importance of a
TLS in the fast developing area of quantum information processing is
in its controlled manipulation as the elementary building block, or
called a  qubit \cite{Oliver}. The controllable dynamics of a driven TLS is at the core of many vastly different state-of-the-art technologies, especially,
solid state implementations of individually addressable
TLSs \cite{Hanson}.
The studies of driven TLSs have quite a rich history, and wide
application for both experimental and theoretical investigations
\cite{Shirley,
 Yuri,kaya,kaya2,Hijii,Greenberg1,Greenberg2,Son09,Ho09,DSS}. Recently,
great progress has been made experimentally using superconducting
devices based on Josephson tunnel
junctions \cite{Tsai,Il,runder,clake,Hof-wang,sun}, optically and
electrically controlled single spins in quantum
dots \cite{Berezovsky,Koppens,Nowack,xuSC,xuNP}, individual charge in
quantum dots\cite{hayashi,Gorman}, and nitrogen vacancy center in
diamond \cite{Jelezko,Fuchs,Buckley}, for the implementation of the
controllable coherent dynamics of qubits.

Periodically time-dependent driving
fields are widely employed to realize the control and manipulations of
qubits. The prototype here described by the semiclassical Rabi model
in the tunneling or localized representation is
\begin{eqnarray}\label{rabi}
    H(t)&=& -\frac{\Delta}{2}\sigma_x - \frac{\varepsilon(t)}{2}\sigma_z  \nonumber\\
    &=&-\frac{\Delta}{2}\sigma_x-\left(\frac{\epsilon}{2}\sigma_z + \frac{A\cos(\omega t)}{2}\sigma_z\right),
\end{eqnarray}
where $\Delta$ is a time-independent tunneling strength,
$\varepsilon(t)$ is a driving field with a static bias $\epsilon$,
$A\cos(\omega t)$ is a periodical driving field with amplitude $A$
and frequency $\omega$, and the symbols $\sigma_x$, $\sigma_y$ and
$\sigma_z$ are the usual Pauli matrices.
We set throughout this paper $\hbar=1$.
Subjecting the Hamiltonian to a rotation about the $y$ axis, we get a
new representation $\exp{(i \pi \sigma_y/4) }H(t) \exp{(-i \pi
  \sigma_y/4) }= -\frac{\Delta}{2}\sigma_z +
\frac{\varepsilon(t)}{2}\sigma_x$. This is the Hamiltonian typically used in quantum
optics and nuclear magnetic resonance in which $\Delta$ is energy difference between the two levels and the
driving term is responsible for the transitions between them.

The Hamiltonian of Eq.~(\ref{rabi}) can represent, for example, a flux
qubit with a
static bias and a driving field in the persistent current basis.
Although this Hamiltonian looks simple, there
exist a wide variety of interesting dynamical features \cite{GH},
like Rabi oscillations, the invalidity of the RWA \cite{Ashhab},
Bloch-Siegert shifts \cite{Shirley,BS-shift,BS-Exp}, non-linear
phenomena due to level crossings induced quantum interference,
coherent destruction of tunneling (CDT) and the possibility of
chaos \cite{CDT1,CDT2,CDT3,CDT4}.
Thus even for this
simplest driven TLS model with a sinusoidal driving field $A
\cos(\omega t)$, it is
a difficult task to present an analytical and exact solution transparently.
In order to discover the driven
tunneling physics analytically, a number of approximate methods, such
as Rabi-RWA and the RWA in a rotating-frame (RWA-RF) of the
Ref.\cite{Ashhab,Hausinger}, have been developed, even though the
dynamics of the Rabi model can be solved exactly by numerical
methods. We discuss
briefly the conditions to make the Rabi-RWA for the Hamiltonian
of Eq.~(\ref{rabi}) and the dynamics of the driven
TLS using the Rabi-RWA method in Appendix~\ref{sub:Rabi-RWA}.

It is interesting and important to study how the counter-rotating (CR)
terms and static bias together influence the dynamics in a wide range
of parameter space, especially in the regime where
the relevant energy scales are of the same order, i.e., $\epsilon \sim
\Delta \sim A \sim \omega$. When the energy scales are near the same,
the perturbation based on $\Delta$, $A$ or $\epsilon$ becomes
invalid. Therefore, it is necessary to develop an alternative
analytical method to extract the physics in this specific parameter regime.
Moreover, only for the zero-bias case and when the CR terms are
dropped (in other
words, the Rabi-RWA method is applicable), the dynamics of
Eq.~(\ref{rabi}) can be solved analytically.
For the finite bias case, the Rabi-RWA Hamiltonian reduced from the
Hamiltonian of Eq.~(\ref{rabi}) by neglecting all the fast oscillatory terms
in the energy eigenbasis can then be analytically solvable. This raises the
question on the validity of the results in the strong driving strength
and off-resonance ($\omega \neq \Xi_0=\sqrt{\Delta^2+\epsilon^2}$)
regimes where the breakdown
of the reduced Rabi-RWA Hamiltonian occurs
\cite{Shirley}.
Recently, it has been found that the effects of CR
terms are significant in different interesting topics, such as quantum
Zeno effect \cite{Zeno}, entanglement evolution \cite{Ficek},
resonance fluorescence \cite{Yan13}, and so on. On the other hand, in
the context of superconducting flux qubits, recent investigations of
strong driving-induced effects on Rabi oscillations
\cite{BS-Exp,Yoshihara} prompt us to
investigate the problem of
how much the Rabi frequency and Bloch-Siegert shift would
change as a function of the static bias $\epsilon$.

The main purpose
of this paper is to demonstrate the significant role of the bias and the
coupling on the time evolution and physical properties of
the driven TLS.
The dynamics of the TLS with far off-resonance and strong driving
strength conditions is interesting but difficult to study
analytically due to its complexities \cite{Hausinger,GH}. The
availability of accurate and
transparent analytic evaluation of the dynamics is very useful and
important. In our
previous papers \cite{ZG2012,Yan15}, we proposed the
counter-rotating-hybridized rotating
wave (CHRW) method to analytically treat the driven dynamics and
numerically calculate Bloch-Siegert shift at zero bias.
In the present paper, we develop a novel analytical method using
the idea of CHRW to systematically study finite bias cases which are realistically operated
in current experiments.
The major difference from our previous papers and the main result of
the current paper are as follows: (i) from the
viewpoint of methodology, in order to take into account the resultant
effects of static bias, our present approach adopts a novel unitary
transformation with a two-parameter ($\xi$ and $\zeta$) generator of
Eq.~(\ref{eq:1})
in contrast with a simple unitary transformation with a single
parameter $\xi$ for the unbiased case in our previous papers
\cite{ZG2012,Yan15}. The present developed CHRW approach paves a way
to investigate the physics of the bias in a driven TLS; (ii) the bias
$\epsilon$ modulates the energy levels of the TLS and therefore
modulates the resonant condition between a driving field and the
TLS. It is interesting and important to investigate the effects of bias
on the dynamics of the TLS and the competition between the bias and
driving. In broad regions of the $\epsilon$-extended parameter space (near
resonance $\Xi_0 \sim \omega$ and off-resonance $\Xi_0 < \omega$), we
find that our CHRW results are in very good agreement with the exact
numerical results. We
demonstrate the significant role of the CR coupling and the bias on the time evolution and physical properties of the driven TLS;
(iii) we calculate the generalized Rabi frequency and the bias-modulated Bloch-Siegert shift that can be measured in experiments
\cite{BS-Exp,Yoshihara}. Our results are in good agreement with those of the flux qubit data presented in Ref.~\cite{Yoshihara};
(iv) after comparing the results obtained by our method with those of the other
RWA methods, the
second order Van Vleck (2nd-VV) perturbation method and the exact numerical method, we demonstrate the
various parameter regimes for the validity of the different methods
and show clearly that our CHRW method is more efficient and accurate than the
other RWA and perturbation methods we discuss.

As the Hamiltonian Eq.~(\ref{rabi}) can be numerically solved easily and quickly
by the Floquet theory, why do we pursue an approximate analytical
solution?  The reasons are as follows.
A good approximate analytical solution should be as simple as
possible, especially mathematically, so that it can be
straightforwardly extended to investigate  more complicated situations where a
numerical study is hard to performed.
More importantly, the main physics should be described pretty well and
the dynamics should be as accurate as possible when compared with the
numerically exact results, at least for
the interesting and concerned range of the parameters (especially
those of experimental relevant parameter regimes).
The simple analytical CHRW method we provide does have such merits.
The CHRW method is beyond the Rabi-RWA and enables one to understand
the physics, such as effects of the CR coupling and the bias,  more clearly.
The important physical properties in the driven system, such as the generalized Rabi frequencies and Bloch-Siegert
shifts which are not easy to extract directly from the time evolution by the numerical
methods, can be calculated in analytical forms.
Our CHRW method can also explore explicitly important physical phenomena, for
example, the CDT, and describe other important dynamical features and behaviors very well for
a wide range of parameters. Furthermore, it can be applied to other more
complicated models or driven open quantum systems where numerically exact
solutions are hard to obtain. Besides, it is interesting to discuss the validity of different RWA
schemes and show how the previous results appear in the various limits of the CHRW method.

The structure of this article is as follows. In Sec.~\ref{intro}, we introduce the driving
TLS model with a static bias.
In Sec.~\ref{sec.CHRWM}, we develop a simple and efficient method to
analytically and quantitatively solve the monochromatically driven
dynamics. In Sec. \ref{sec.result}, we analyze the
dynamics in a wide parameter regime including the cases of resonance, near
resonance and far off resonance. Moreover, we demonstrate in
Sec.~\ref{sec.Rabi} the effects
of the CR wave
terms and the bias on the dynamics, the generalized Rabi frequency and the
Bloch-Siegert shift. Finally, we give
the parameter regions for the CHRW method to be valid before we present a short conclusion in Sec.~\ref{sec.sum}.

\section{COUNTER-ROTATING HYBRIDIZED ROTATING WAVE METHOD} \label{sec.CHRWM}

In this section, we describe the CHRW method
to calculate the
driven tunneling dynamics for the finite bias case \cite{ZG2012,Yan15,ZL}.
In the CHRW method a time-dependent unitary transformation $\exp(S)$ is applied
to the system and
we propose the generator of the unitary transformation to be
\begin{equation}
  \label{eq:1}
S=-i\frac{A}{2\omega}\sin(\omega t)[\xi\sigma_z+\zeta\sigma_x ].
\end{equation}
The two significant parameters $\xi$ and $\zeta$ introduced in $S$
will be determined later on. We use the time-dependent Schr\"odinger
equation $ i\frac{d }{dt}|\Psi(t)\rangle = H(t) |\Psi(t)\rangle$  to solve the
dynamics. After the unitary transformation, we obtain readily the
interaction picture formulas with $|\Psi'(t)\rangle=\exp(S)|\Psi(t)\rangle$ and
$i\frac{d }{dt}|\Psi'(t)\rangle = H'(t) |\Psi'(t)\rangle$. Here the
transformed Hamiltonian is
\begin{eqnarray} \label{H'}
  H' &=& -\frac{\Delta}{2}\left[\sigma_x -\frac{1-\cos \Theta}{X^2} \xi \left(\xi \sigma_x-\zeta\sigma_z\right)+ \frac{\sin \Theta}{X}\xi \sigma_y \right] \\ \nonumber
  &&-\frac{\varepsilon(t)}{2}\left[\sigma_z +\frac{1-\cos\Theta}{X^2} \zeta \left(\xi \sigma_x-\zeta\sigma_z\right) - \frac{\sin\Theta}{X}\zeta \sigma_y \right] \\ \nonumber
   && +\frac{A}{2}\left(\xi\sigma_z+\zeta\sigma_x\right)\cos(\omega
   t),
\end{eqnarray}
where
\begin{eqnarray}
 \Theta&=&\frac{A}{\omega}X\sin(\omega t),
\label{eq:Theta} \\
 Z &=& \frac{A}{\omega}X, \label{eq:Z}\\
 X&=&\sqrt{\xi^2+\zeta^2}.
\label{eq:X}
\end{eqnarray}
After making use of the relation \cite{Ashhab}
\begin{eqnarray}
  \exp(iz \sin\alpha) &=& \sum^{\infty}_{n=-\infty}J_n(z) e^{in \alpha},
\end{eqnarray}
where $J_n(z)$ are the nth-order Bessel functions of the first kind,
we divide the Hamiltonian into three parts $H'=H'_0+H'_1+H'_2$
according to the order of the harmonics (photon transfer process: $0$
photon, $1$ photon,$...$, $m$ photons), where
\begin{eqnarray} \label{H0} \label{H1}
  H_0'&=& -\frac{\tilde{\Delta}}{2}\sigma_x -\frac{\tilde{\epsilon}}{2} \sigma_z, \\
  H_1'&=& - \frac{(\Delta\xi-\epsilon\zeta)}{X}J_1\left(Z\right)\sin(\omega t)\sigma_y  \nonumber \\
  &&- \frac{A}{2}\left[1-\xi- \zeta^2 J_c \right]\cos(\omega t)\sigma_z   \nonumber \\
  &&+ \frac{A}{2}\zeta \left[1- \xi J_c\right]\cos(\omega t)\sigma_x,  \\
  H_2'&=& ~\frac{A}{2}\frac{\zeta}{X}J_1\left(Z\right)\sin(2\omega t)\sigma_y  \nonumber \\
  &&- \frac{(\Delta\xi-\epsilon\zeta)}{X^2} J_2\left(Z\right) \cos(2\omega t)(\xi\sigma_x-\zeta\sigma_z) \nonumber \\
  &&+ \frac{A}{2} \frac{\zeta}{X^2} J_2\left(Z\right)\cos(3\omega t)(\xi\sigma_x-\zeta\sigma_z) \nonumber \\
  &&- \frac{(\Delta\xi-\varepsilon(t)\zeta)}{X^2}\sum_{n=2}^{\infty}\left\{ X J_{2n-1}(Z) \sin[(2n-1)\omega t] \sigma_y \right. \nonumber \\
  &&+ \left. J_{2n}(Z)\cos(2n\omega t)(\xi\sigma_x-\zeta\sigma_z)\right\},
\end{eqnarray}
and the parameters  $\tilde{\Delta}$, $\tilde{\epsilon}$ and $J_c$ are
defined as
\begin{eqnarray}
  \tilde{\Delta}&=&\Delta-\frac{\xi}{X^2}\left[ 1-J_0\left(Z\right)\right](\Delta\xi-\epsilon\zeta),  \label{Delta_ren}\\
  \tilde{\epsilon}&=& \epsilon ~+\frac{\zeta}{X^2}\left[ 1-J_0\left(Z\right)\right](\Delta\xi-\epsilon\zeta),  \label{epsilon_ren} \\
  J_c &=& \frac{1-J_0\left(Z\right)-J_2\left(Z\right) }{X^2}.
\end{eqnarray}
Note that the zero-$\omega$ Hamiltonian
$H'_0$  consists of the renormalized tunneling $\tilde{\Delta}$
and the renormalized bias $ \tilde{\epsilon}$, both
with a modified factor $J_0(A X/\omega)$ including infinite order
of $A$ [see Eqs.~(\ref{Delta_ren}) and (\ref{epsilon_ren})].
$H'_1$ contains all single-$\omega$ terms in the transformed
Hamiltonian with a factor $J_1(A X/\omega)$ which relates to single-photon assisted
transitions and $H_2'$ includes all higher order harmonic terms
such as $\cos(n\omega t)$ and $\sin(n\omega t)$ with $n \ge 2$.

The Hamiltonian $H_0'$ can be diagonalized by a unitary matrix $U = u
\sigma_z - v\sigma_x$ with
\begin{eqnarray}
u=\sqrt{ {\frac{1}{2}}-\frac{\tilde \epsilon }{2\widetilde\Xi}},
~~~ v=\sqrt{ {\frac{1}{2}}+\frac{\tilde \epsilon }{2\widetilde\Xi}},
\end{eqnarray}
into the form
\begin{eqnarray} \label{H0p}
\tilde{H}_0=\frac{\widetilde\Xi}{2} \tau_z,
\end{eqnarray}
where $\tau_z$ is the $z$-component spin operator in the energy
eigenbasis and
\begin{equation}
  \label{eq:Xi_ren}
\widetilde\Xi=\sqrt{\tilde{\Delta}^2+\tilde{\epsilon}^2}
\end{equation}
is the renormalized energy splitting.  At the same time $H_1'$ becomes a
little bit complicated $\tilde{H}_1=U^{\dag} H_1' U$ in the new energy basis,
\begin{eqnarray} \label{H1p}
  \tilde H_1&=&  \frac{(\Delta\xi-\epsilon\zeta)}{X}J_1\left(Z\right)\sin(\omega t)\tau_y  \nonumber \\
  &&+ \frac{A}{2}\left[1-\xi- \zeta^2 J_c \right]\cos(\omega t)\left(\frac{\tilde \epsilon}{\widetilde{\Xi}}\tau_z + \frac{\tilde \Delta}{\widetilde{\Xi}}\tau_x\right)   \nonumber \\
  &&+ \frac{A}{2}\zeta \left[1- \xi J_c\right]\cos(\omega
  t)\left(\frac{\tilde \epsilon}{\widetilde{\Xi}}\tau_x - \frac{\tilde
      \Delta}{\widetilde{\Xi}}\tau_z\right),
\end{eqnarray}
where $\tau_x$ and $\tau_y$ are, respectively, the $x$-component and
the $y$-component spin operators in the energy eigenbasis.
Then, in order to make the driving interaction term
$\tilde{H}_1=U^{\dag} H_1' U$ hold the RWA-like interaction form, we
choose the two proper parameters $\xi$ and $\zeta$ to satisfy the
following two self-consistent conditions. First, we require the
coefficient of counter-rotating terms $\exp(\pm i\omega t) (\tau_x
\pm i\tau_y)/2$ in Eq.~(\ref{H1p}) to vanish, so we have
\begin{eqnarray} \label{xi}
  0 &=&  \frac{A}{2}\left[\frac{\tilde{\Delta}}{\widetilde{\Xi}}\left[1-\xi-\zeta^2 J_c \right]
+ \frac{\tilde{\epsilon}}{\widetilde{\Xi}} \zeta \left(1- \xi J_c \right)\right]-\frac{\Delta\xi-\epsilon\zeta}{X} J_1\left(Z\right).
\end{eqnarray}
Second, we require the coefficients of $ \frac{A}{2}\cos(\omega
t)\tau_z$ term in $H_1'$ to be zero. This then leads to
\begin{eqnarray} \label{zeta}
0 &=& \tilde{\epsilon}\left(1-\xi -\zeta^2 J_c\right) -\tilde{\Delta}\zeta\left(1-\xi J_c\right).
\end{eqnarray}
The two parameters $\xi$ and $\zeta$ can be self-consistently solved
by Eqs.~(\ref{xi}) and (\ref{zeta}). Notice that from Eq.~(\ref{H'})
to Eq.(\ref{zeta}), no approximation is involved.

In the following
treatment, we keep all the $0th$ and $1st$ harmonics of the
transformed driven Hamiltonian ($n \omega$, $ n=0, 1$)
$\tilde{H}_0+\tilde{H}_1$ and neglect the
higher-order harmonic terms of $\tilde{H}_2=U^{\dag}H_2' U$
that involves all
multi-$\omega$ or  multi-photon assisted transitions
($n \omega$, $n=2,3,4...$) in the transformed energy eigenbasis.
The validity of the omission of $\tilde{H}_2$ or
$H'_2$ depends on the effects of the higher-frequency driving terms ($n
\ge 2$), i.e. the fast-oscillating term, generally accompanying the
second-order or higher-order Bessel functions.
Its contribution to the dynamics is not prominent except for the
ultra-strong driving strength case.
This is called the CHRW method
and we
obtain the reformulated RWA Hamiltonian
\begin{eqnarray}\label{CHRW}
 H_{\mathrm{CHRW}} &=& \tilde{H}_0+\tilde{H}_1=\frac{\widetilde{\Xi}}{2}\tau_z +  \frac{\tilde{A}}{2}(\tau_{+}\exp(-i \omega t) + \tau_{-} \exp(i \omega t)),
\end{eqnarray}
where $\tau_{\pm}=(\tau_{x}\pm i\tau_{y})/2$, $\widetilde{\Xi}$ is the
renormalized energy splitting involved with the static bias's
modulation, and $\tilde{A}$ is the renormalized amplitude of the
driving field resulting from the combination of the CR coupling and
static bias,
\begin{eqnarray}
  \tilde{A}
  &=& A
  \left[\frac{\tilde{\Delta}}{\widetilde{\Xi}}\left(1-\xi-\zeta^2 J_c
    \right)
+\frac{\tilde{\epsilon}}{\widetilde{\Xi}} \zeta \left(1- \xi J_c
\right)\right] \\
&=&\frac{\Delta\xi-\epsilon\zeta}{X} 2 J_1\left(\frac{A}{\omega}
  X\right).
\label{A_ren}
\end{eqnarray}
In obtaining the second equality of Eq.~(\ref{A_ren}), we have used
Eq.~(\ref{xi}).

One can see that the effects of the driving and bias have been taken
into account in our treatment which leads to the renormalization of
the significant physical properties. An interesting and key point
about the CHRW method is that the CHRW Hamiltonian, Eq.~(\ref{CHRW}),
has the same
mathematical formulation as the Rabi-RWA one except for the
renormalized physical quantities. Therefore,
it is mathematically straightforward to obtain the CHRW dynamics
by the well-known Rabi-RWA one.
We note here that the CHRW method that neglects the
higher-order harmonic terms of $\tilde{H}_2$ with $n\ge 2$
is not a perturbation based on small
tunneling, bias or driving strength. In principle, when the
driving frequency is greater than the energy splitting, i.e. $\omega
\ge \Xi_0=\sqrt{\Delta^2+\epsilon^2}$, the processes involving zero
and single photon are dominant. In this case, one can safely
neglect the contributions from the higher-order harmonic terms in the
transformed frame.
When $\omega < \Xi_0=\sqrt{\Delta^2+\epsilon^2}$ and $A \le \omega$,
neglecting the high harmonic terms in the
transformed Hamiltonian still yields pretty good results.
We will verify these in Sec.~\ref{sec.sum} by detailed
examination of the dynamics and the general Rabi frequency of the CHRW
method with the exact numerical results.
It is only when $\omega \ll  \Xi_0=\sqrt{\Delta^2+\epsilon^2}$ and
in the very strong driving case that the multi-photon processes might
make considerable contributions to the dynamics and the physical
quantities, and in this case the higher harmonic terms in  $\tilde{H}_2$
can not be completely omitted.
Thus our CHRW method is a reliable and effective approach to
investigate the bias-modulated Rabi model in a wide range of parameters.

In the following, we calculate an important property in the driven
dynamics, namely the occupation probability $P_{\mathrm
  {up}}(t)$ \cite{Ashhab} in the CHRW method.
$P_{\mathrm {up}}(t)$ denotes the probability
for the system at time $t$ to be in the spin-up
state of the $\sigma_z$ operator in the original spin
basis of Eq.~(\ref{rabi}) while it is initially in the spin-down
state of the same $\sigma_z$ operator
(i.e., $P_{\mathrm  {up}}(0)=0$).
Because the generator $S$ is a function of $\sin(\omega t)$, the
initial system state is
the same as that in the unitarily transformed frame,
namely, $|\Psi'(0)\rangle=|\Psi(0)\rangle$. Since we have also used a
unitary matrix $U$ to obtain $H_{\rm CHRW}$,
the corresponding system states
 $|\tilde{\Psi}(t)\rangle=U^{\dag}|\Psi'(t)\rangle$
satisfies the Schr\"odinger equation,
\begin{equation}\label{wave}
    i\frac{d}{dt} |\tilde{\Psi}(t)\rangle = H_{\rm CHRW} |\tilde{\Psi}(t)\rangle.
\end{equation}
Let us write  $|\tilde{\Psi}(t)\rangle$,
in terms of the eigenstates of the $\tau_z$ operator as
$|\tilde{\Psi}(t)\rangle=c_1(t) |s_{1}\rangle +c_2(t) |s_{2}\rangle$ with $\tau_z |s_{1}\rangle=-|s_{1}\rangle$ and $\tau_z |s_{2}\rangle= |s_{2}\rangle$.
Substituting it back to the Schr\"odinger equation (\ref{wave}),
we can readily solve for $c_{1}(t)$ and  $c_{2}(t)$ for the initial
condition of the TLS being in the spin-down state of $\sigma_z$, i.e.,
$\langle\sigma_z(0)\rangle=-1$ which corresponds to
$c_1(0)=-u, c_2(0)=-v$, as \cite{scu}:
\begin{eqnarray}
  c_{1}(t) &=&   e^{ i\frac{\omega t}{2}}
  \left\{-u\left[\cos\left(\frac{{\Omega}_{\rm R} t}{2}\right)+i
      \frac{{\tilde\delta}}{{\Omega}_{\rm R}} \sin\left(\frac{{\Omega}_{\rm R}
          t}{2}\right) \right] +i v \frac{\tilde{A}}{{\Omega}_{\rm R}}
    \sin\left(\frac{{\Omega}_{\rm R} t}{2}\right) \right\},
\label{eq:c1}\\
  c_{2}(t) &=&  e^{ -i\frac{\omega t}{2}}
  \left\{-v\left[\cos\left(\frac{{\Omega}_{\rm R} t}{2}\right)-i
      \frac{{\tilde\delta}}{{\Omega}_{\rm R}} \sin\left(\frac{{\Omega}_{\rm R}
          t}{2}\right) \right] +i u \frac{\tilde{A}}{{\Omega}_{\rm R}}
    \sin\left(\frac{{\Omega}_{\rm R} t}{2}\right) \right\},
\label{eq:c2}
\end{eqnarray}
where
\begin{eqnarray}
\Omega_{\rm R}&=&\sqrt{\tilde{\delta}^2+\tilde{A}^2},
\label{eq:Rabi_freq}\\
\tilde{\delta}&=&\widetilde{\Xi}-\omega
\label{eq:ren_detuning}
\end{eqnarray}
are, respectively, the modulated Rabi frequency and the
renormalized detuning parameter of the CHRW method.
Thus the population of the spin-up state $\left(
\begin{array}{c}
1 \\
0
\end{array}
\right)$ in the $\sigma_z$ basis at time $t$ for initial
$P_{\mathrm  {up}}(0)=0$ is $P_{\mathrm {up}}^{\mathrm{CHRW}}(t)= \langle \Psi(t)| \frac{\sigma_z +1}{2} |\Psi(t)\rangle$, in which
\begin{eqnarray}\label{pt-CHRW}
\langle \Psi(t)| \sigma_z |\Psi(t)\rangle &=& \langle \tilde{\Psi}(t)| U^{\dag} e^{S(t)}\sigma_z e^{-S(t)} U|\tilde{\Psi}(t)\rangle \\ \nonumber
&=& \left\{1-\frac{\zeta^2}{X^2}\left[1-\cos\Theta(t) \right]\right\}\left\{(v^2-u^2)\left(|c_1|^2-|c_2|^2\right)-2uv(c_2^{*}c_1+c_1^{*}c_2)\right\} \\ \nonumber
&&-\frac{\xi\zeta}{X^2}[1-\cos\Theta(t)]\left\{2uv\left(|c_1|^2-|c_2|^2\right)-(v^2-u^2)(c_2^{*}c_1+c_1^{*}c_2)\right\} \\ \nonumber
&&-\frac{\zeta}{X}\sin\Theta(t) i \left(c_2^{*}c_1-c_1^{*}c_2\right).
\end{eqnarray}
Note that $c_1$ and $c_2$ are time-dependent and their expressions
are given, respectively, in Eqs.~(\ref{eq:c1}) and (\ref{eq:c2}), and the parameters
$\Theta$ and $X$ are defined, respectively, in  Eqs.~(\ref{eq:Theta})
and (\ref{eq:X}).
The general renormalized Rabi frequency ${\Omega}_{\rm R}$ of
Eq.~(\ref{eq:Rabi_freq}) has taken into account the effects of CR
terms and static bias on frequency shifts and will give the
Bloch-Siegert shift in a simple way (will be described
in Sec.~\ref{sec.Rabi}).
Physically, the renormalized quantities in the transformed Hamiltonian
Eq.~(\ref{CHRW}) can be
detected from the general Rabi frequency and the
Bloch-Siegert shift.

As discussed in our previous work \cite{ZG2012},  we demonstrated
clearly that the result of the RWA-RF method \cite{Hausinger} is a
limiting case of the CHRW method for the zero bias case.
For the resonance
condition $n\omega+\epsilon=0$ to hold, only one value of $n$ is
kept in the RWA-RF approach. One usually identifies the kind of resonance with a given value
of $n$ as an $n$-photon process. For example, the Rabi-RF
resonance condition for the case $n=-1$ means $\epsilon=\omega$ (the
difference from the traditional condition
$\omega=\Xi_0=\sqrt{\Delta^2+\epsilon^2}$ in the Rabi-RWA case will become clear
shortly). Therefore, the effective RWA-RF Hamiltonian is written as
\begin{equation}\label{rwa-rf}
    H_{\mathrm{RWA-RF}}= - J_n\left(\frac{A}{\omega}\right) \frac{\Delta }{2}\sigma_x .
\end{equation}
Thus, the probability $P_{\mathrm {up}}(t)$ of the RWA-RF approach in
Ref.~\cite{Ashhab} is obtained as
\be \label{rrwa}
P_{\mathrm {up}}^{\mathrm {RWA-RF}}(t)=\sin^{2}\left\{J_n\left(\frac{A}{\omega}\right)\frac{\Delta t}{2}\right\}
\ee
whose amplitude is always one for any driving parameter. With
parameters satisfying the resonance condition, the oscillation
frequency is $J_n\left(\frac{A}{\omega}\right) \Delta$. This means
that the $P_{\mathrm {up}}(t)$ always exhibits a full periodic
oscillation between the up and down states of $\sigma_z$ except the
CDT case where $P_{\mathrm {up}}^{\mathrm {RWA-RF}}(t)\equiv 0$ due to
$J_n\left(\frac{A}{\omega}\right)=0$. Therefore, the result
of Eq.~(\ref{rrwa}) of the RWA-RF approach
is distinguished from that of the CHRW method
[c.f. Eq.(\ref{pt-CHRW})]. This treatment simply corresponds to the case
$\xi=1$ and $\zeta=0$ of the CHRW method, which is only valid in the
limit of a really strong driving strength ($A \gg \omega, \Delta$) and
with the condition $|n\omega+\epsilon|=0$. Moreover. in Ref. \cite{Hausinger}, the Van
Vleck perturbation theory is used to get the survival probability to
second order in $\Delta$ for a finite static bias.
We will show in the next section that
the CHRW method gives a significantly better description of
the system dynamics
than the previous RWA or perturbative treatments
as compared with the numerically exact results.


\section{Driven Quantum dynamics}
\label{sec.result}
We systematically discuss here the dynamics of the driven TLS with a
bias in different parameter regimes: at resonance, near resonance, and
far-off resonance. With the increase of the driving strength from the weak- to strong-coupling regime, a rich distinct dynamics
can be observed. We also compare all the results of the CHRW approach
with those of the other methods, namely, the Rabi-RWA method, the
RWA-RF method, the 2nd-VV perturbation method
and the numerically exact method. Moreover, we discuss the
results of time evolutions through a frequency spectrum analysis
to show the accuracy of our CHRW method.


\subsection{At resonance and near resonance}

\begin{figure}[htbp]
  \includegraphics[width=8cm]{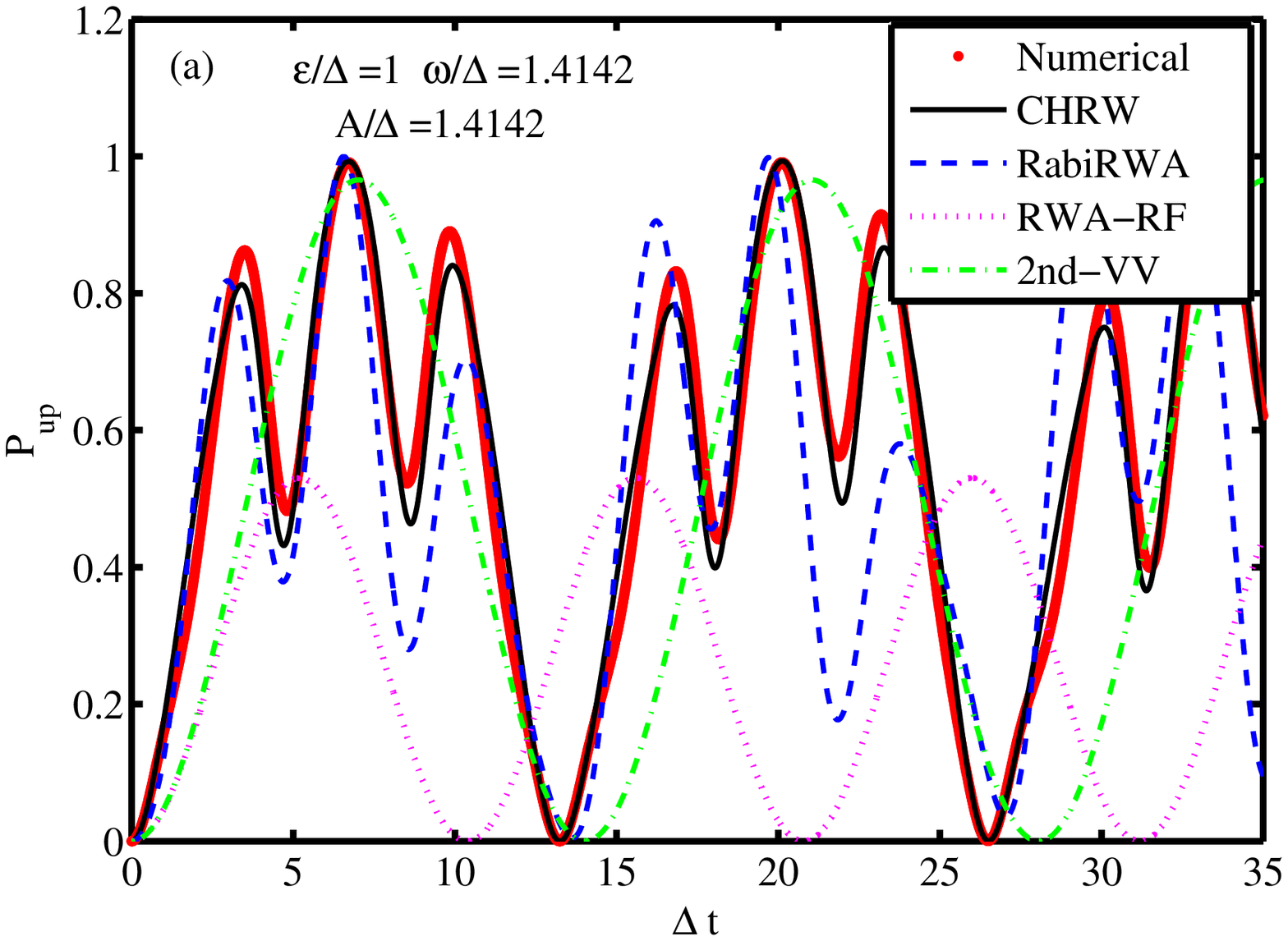}
  \includegraphics[width=8cm]{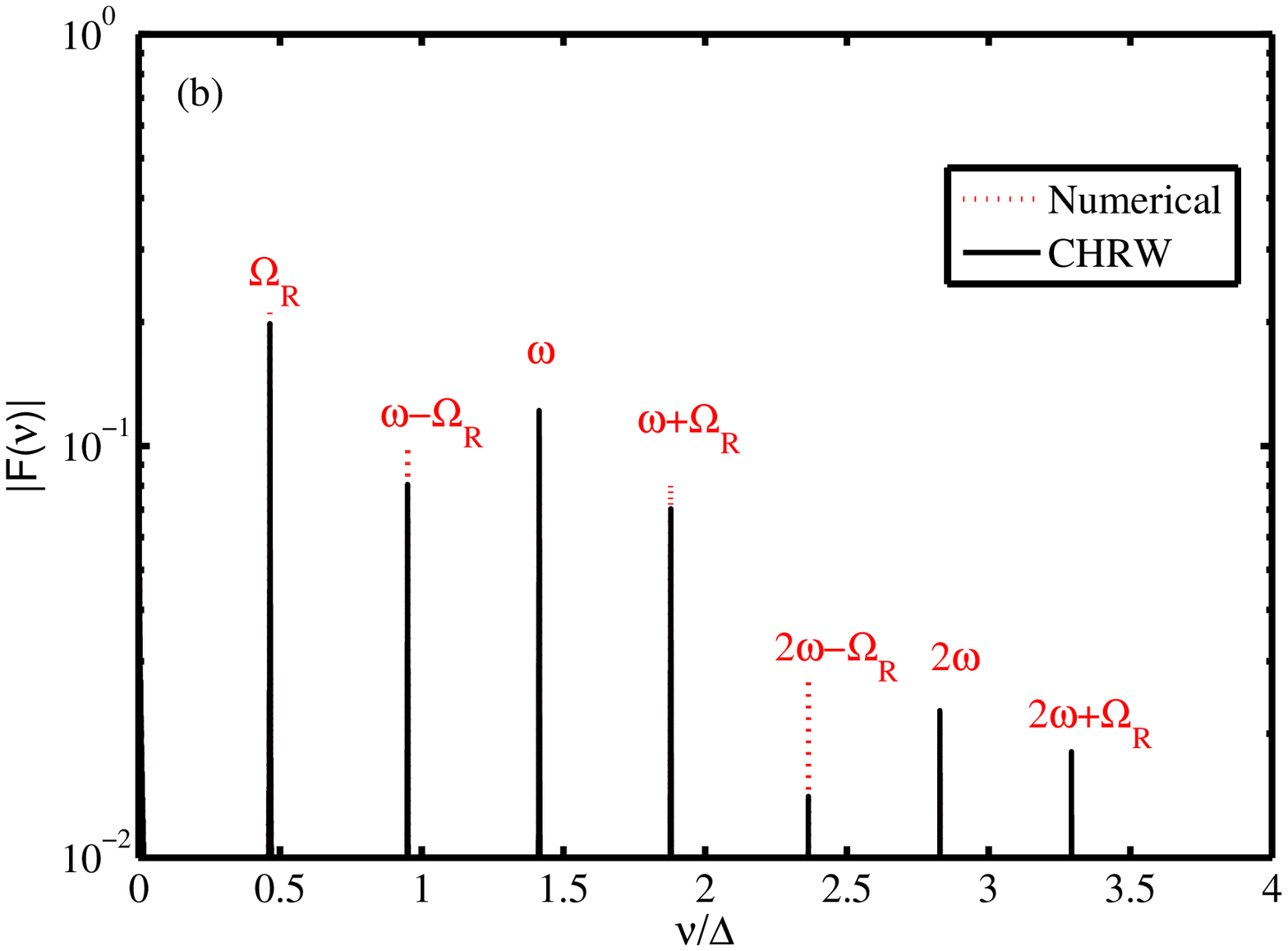}
  \includegraphics[width=8cm]{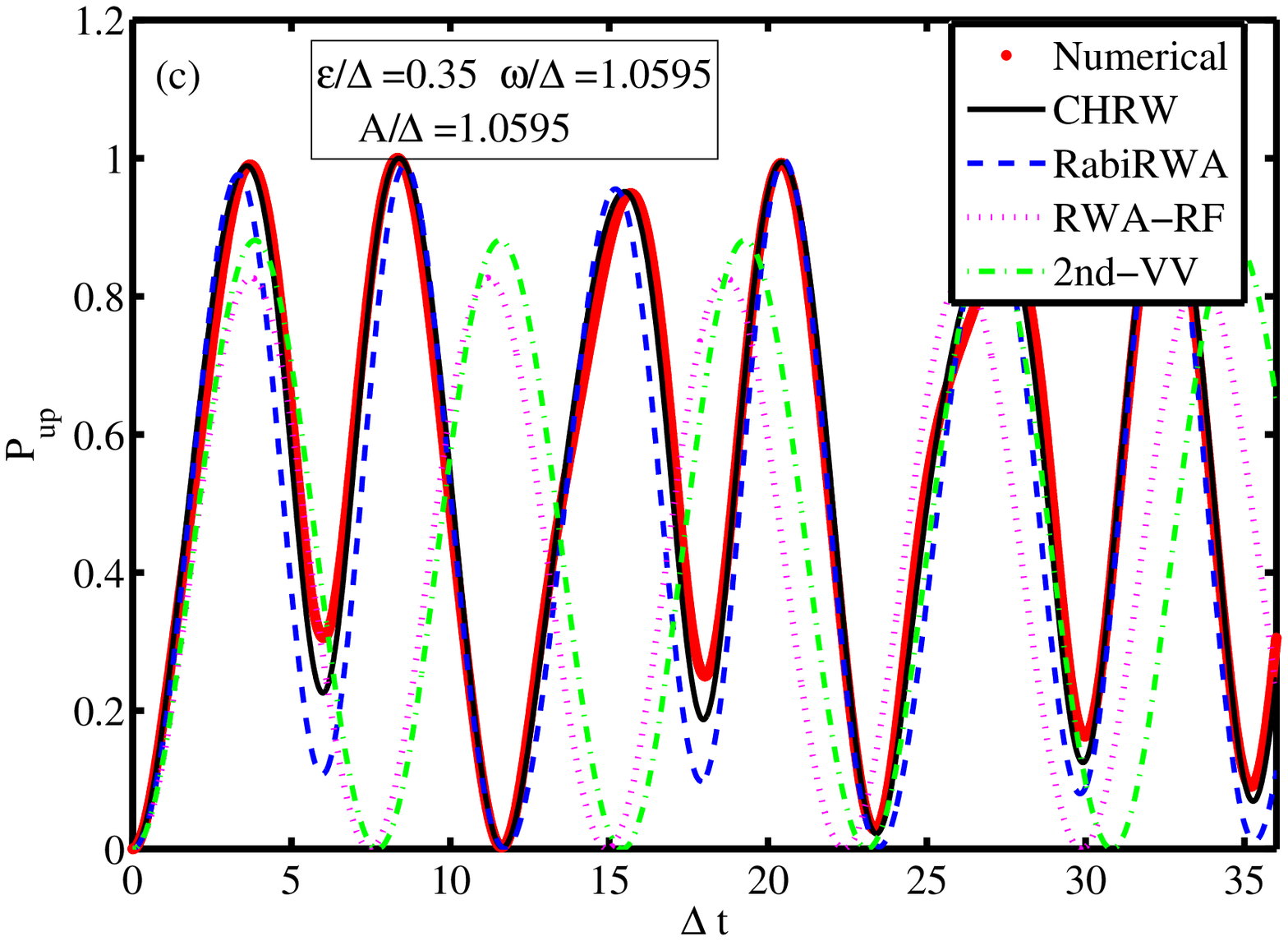}
  \includegraphics[width=8cm]{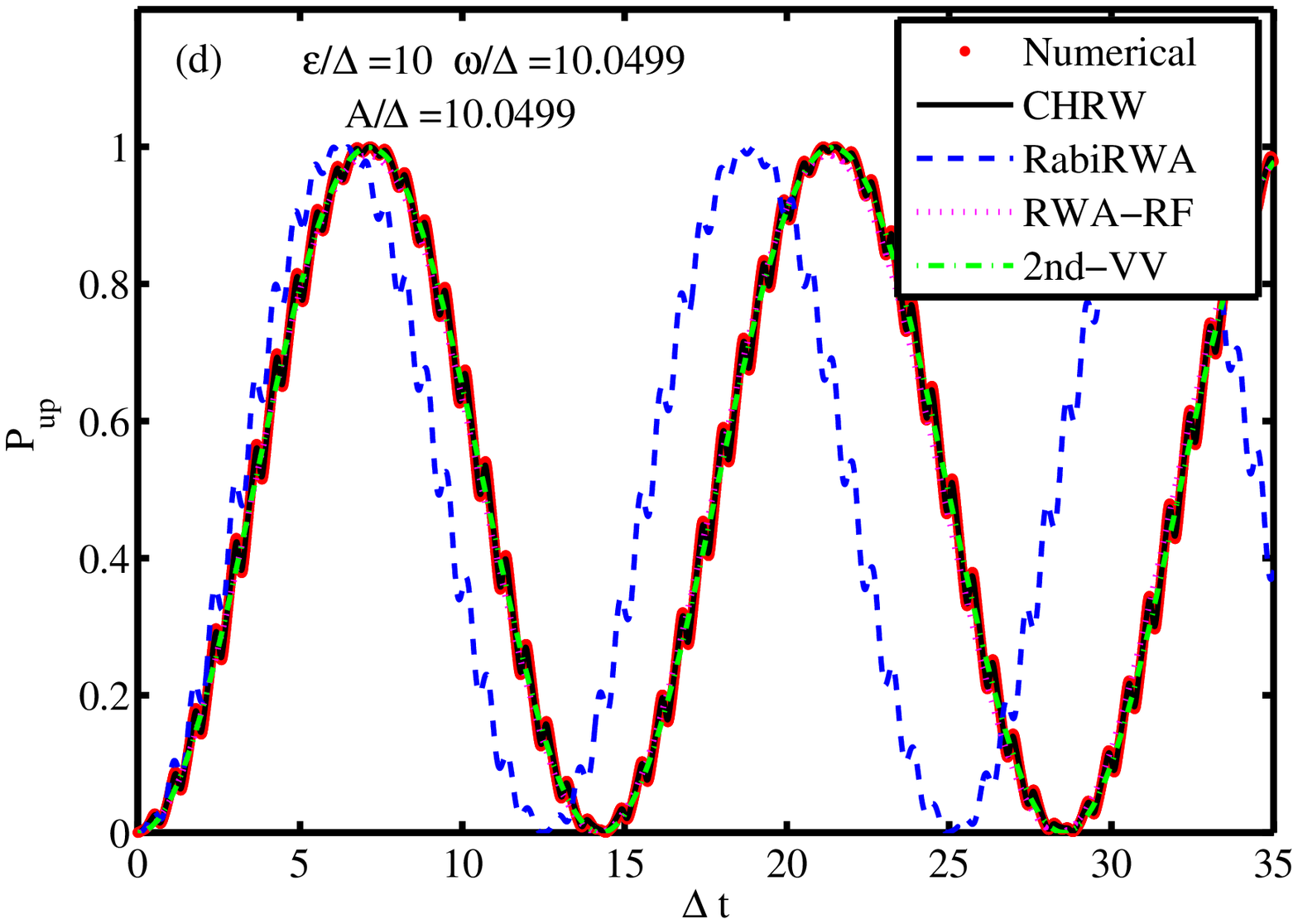}\\
  \caption{(Color online) Time evolutions of $P_{\rm
      up}(t)=\langle\frac{\sigma_z(t)+1}{2}\rangle$ as a function of
    $\Delta t$ for different values of the bias (a) $\epsilon/\Delta=
    1.0$, (b) $0.35$ and (c) $10$ in the on-resonance case
    ($\omega=\Xi_0=\sqrt{\Delta^2+\epsilon^2}$).
The driving strength $A$ is set
    to be $A/\Xi_0=1$. The Fourier transform $F(\nu)$ of $P_{\rm
      up}(t)$ in (a) is shown in (b) with a discrete set of frequency
    components. }\label{fig1}
\end{figure}

The bias modulates the energy levels of the TLS and
therefore modulates the resonant condition between a driving field and
the TLS.
Let us take a look at the dynamics at resonance ($\omega=\Xi_0$) and near
resonance ($\omega\sim\Xi_0$) for the small to large bias cases. In
Fig. {\ref{fig1}}, we show the occupation probability at
resonance with $A/\omega = 1$. For comparison, we also give the
results of the other different approaches.
One can see that
the results of the CHRW are in good agreement with the numerically
exact results. While the Rabi-RWA method works for $\epsilon/\Delta
< 1$ with small deviation in amplitude from the numerically exact
one [see Fig.~\ref{fig1}(c)], the results of the Rabi-RF and 2nd-VV
methods give different frequencies of oscillation from the numerically
exact result. Nevertheless, the Rabi-RF and 2nd-VV could correctly
predict the frequency of main oscillation for $\epsilon/\Delta \gg 1$
 but could not give correctly the small
wiggling amplitudes of the fast oscillations shown in Fig.~\ref{fig1}(d).
When $\epsilon/\Delta = 1$, the population probabilities
$P_{\mathrm{up}}(t)$  obtained by the Rabi-RWA, Rabi-RF and 2nd-VV
methods all show considerable difference from the exact, numerical
result [see Fig.~\ref{fig1}(a)].  But the CHRW result still agrees
rather well with the corresponding numerical result.
In Fig. {\ref{fig1}}(b), we show the Fourier transform of
the $P_{\mathrm{up}}(t)$ in Fig. {\ref{fig1}}(a):
\begin{equation}
  \label{eq:Fnu}
F(\nu)=\int_{-\infty}^{\infty}dt P_{\mathrm{up}}(t)\exp(i \nu t).
\end{equation}
The values (or positions) of the discrete frequencies obtained by the
CHRW method are
precisely the same as those obtained by
the numerically exact method. One can see that there exist two
dominating oscillation frequencies with larger weight, one
corresponding to the driving frequency $\omega/\Delta=\sqrt{2}$ and the
other corresponding to the Rabi frequency
$\Omega_{\rm  R}/\Delta=0.4643$. Moreover,
there exhibit the components of the frequencies $n\omega\pm\Omega_{\rm
  R}$ ($n\geq 1$) and $m\omega$ ($m \geq 2$) with small weight, which
are consistent with the formula in Eq.~(\ref{pt-CHRW}).

\begin{figure}[htbp]
  \includegraphics[width=8cm]{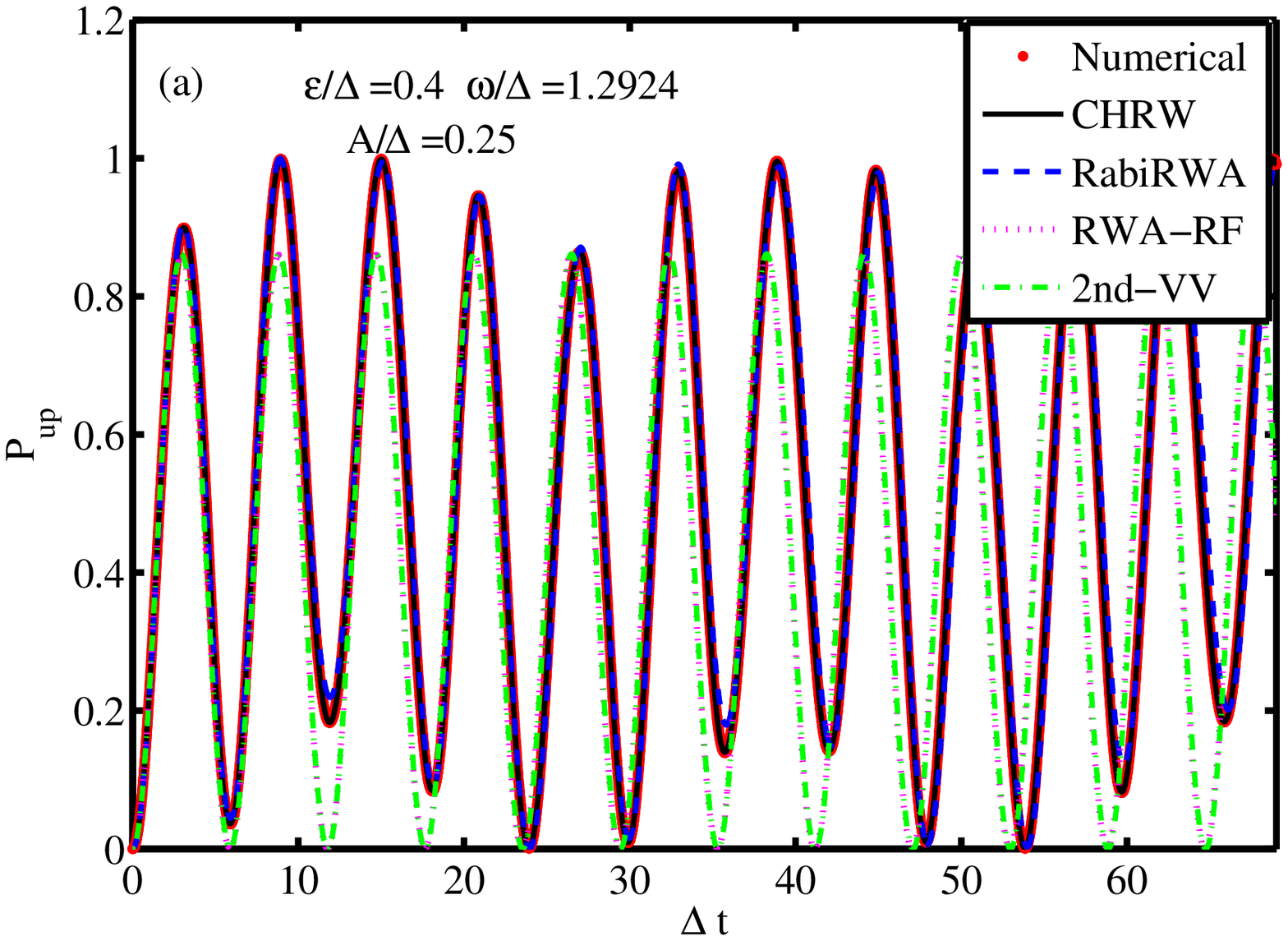}
  \includegraphics[width=8cm]{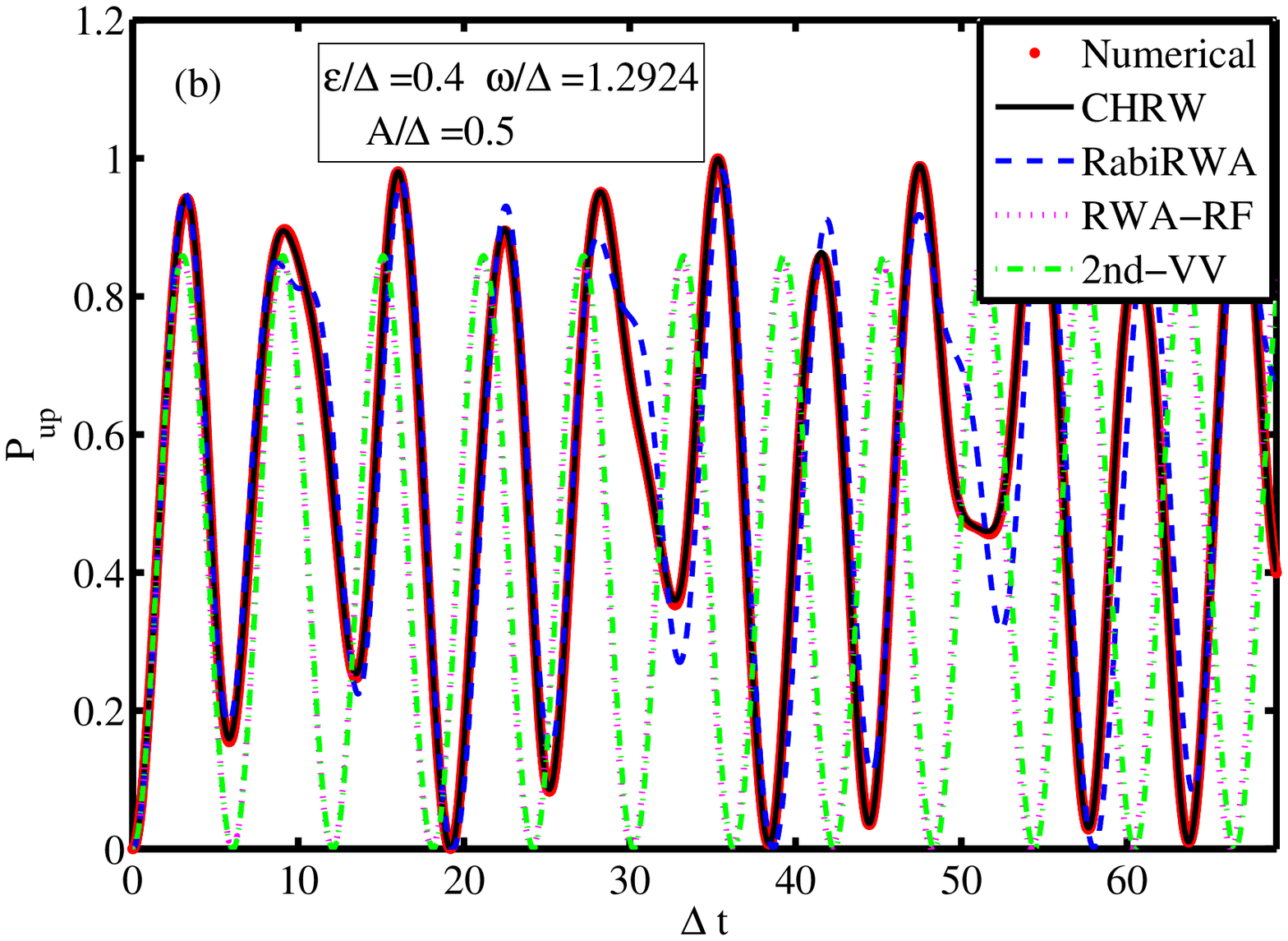}
  \includegraphics[width=8cm]{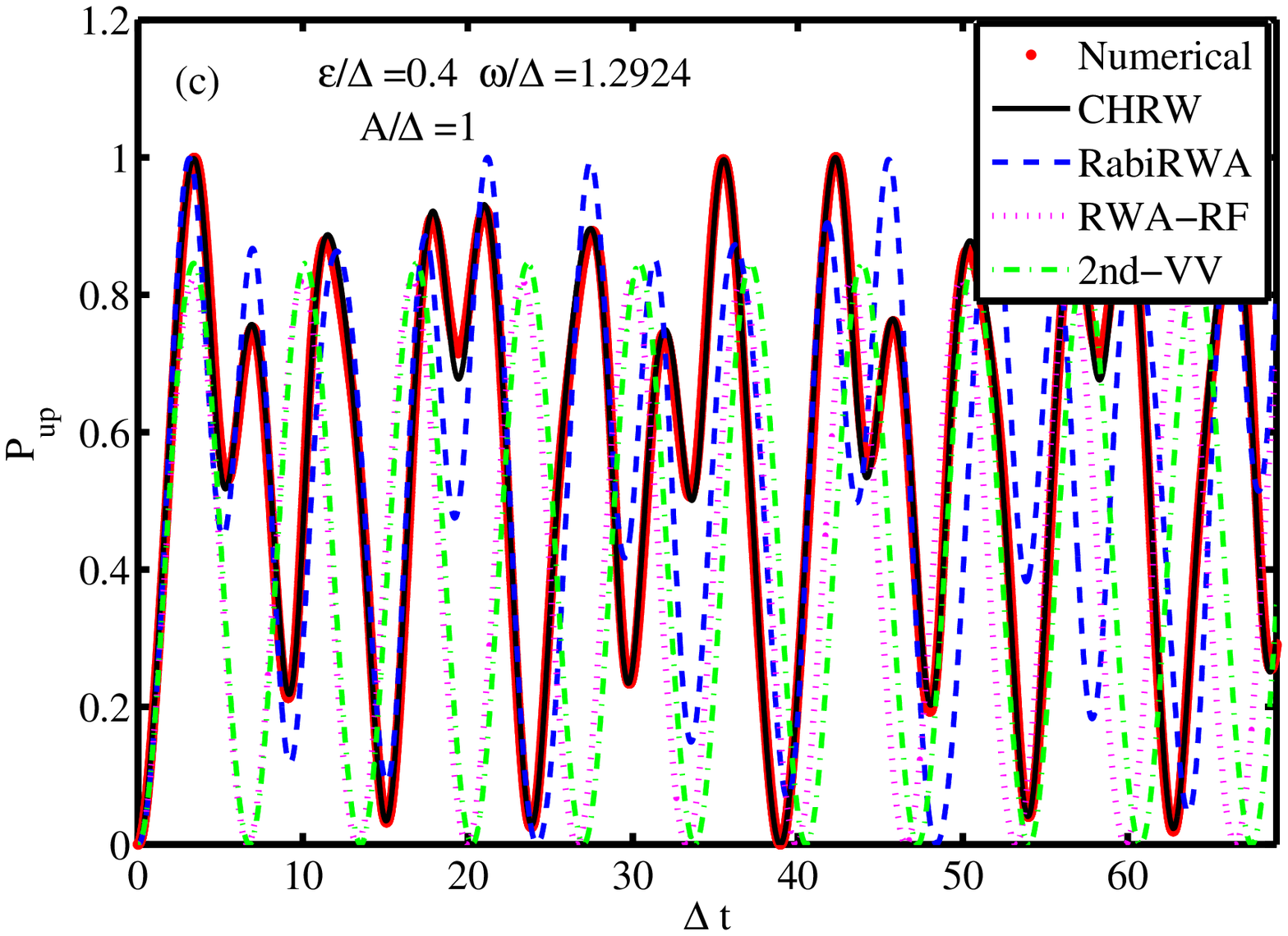}
  \includegraphics[width=8cm]{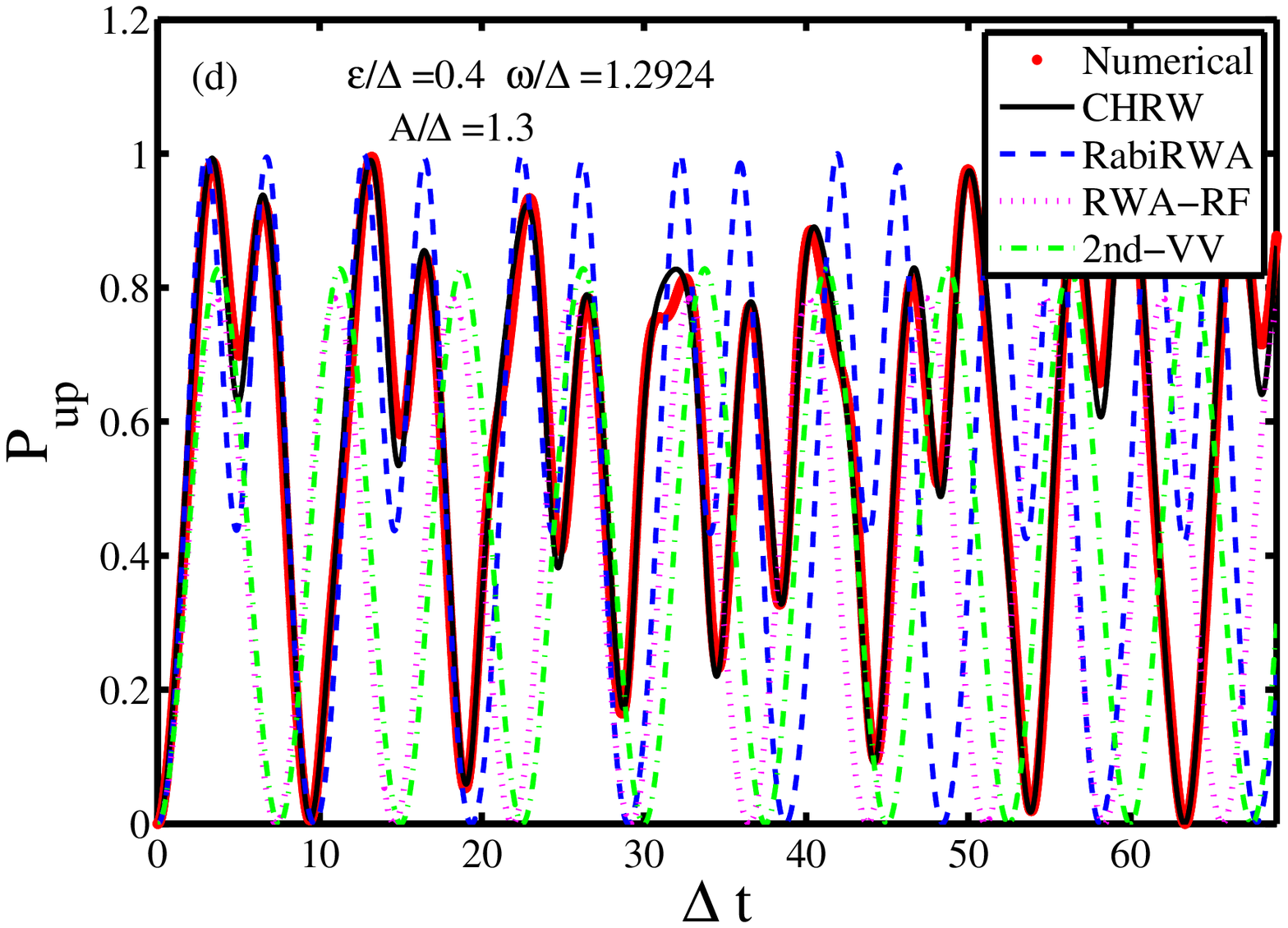}\\
  \caption{(Color online) Time evolutions of $P_{\rm
      up}(t)=\langle\frac{\sigma_z(t)+1}{2}\rangle$ as a function of
    $\Delta t$ for different values of the driving strength
    (a) $A/\Delta= 0.25$, (b) $0.5$, (c) $1.0$ and (d) $1.3$  in the
    near-resonance case ($\omega=1.2\Xi_0=1.2924\Delta$).
The bias $\epsilon$ is set to be a fixed value of
$\epsilon/\Delta=0.4$.}\label{fig2}
\end{figure}

We show the near-resonance dynamics ($\omega=1.2\Xi_0=1.2924\Delta$)
for several moderate driving strengths in Fig.~{\ref{fig2}}.
It is easy to check that for a very weak driving strength the results
of all methods are nearly the same [see Fig.~\ref{fig2}(a)].
However,  the Rabi-RWA method breaks down when $A/\Delta>0.5$ [see
Fig.~\ref{fig2}(c)-(d)]. Meanwhile, the deviation of the RWA-RF and
2nd-VV results from the numerically
exact results becomes much larger with the increase of the driving
strength [see Fig.~\ref{fig2}(b)-(d)].
In contrast, our CHRW method works quite well for all the parameters
used in Fig.~\ref{fig2}.
The time evolutions of $P^{\mathrm{CHRW}}_{\mathrm{up}}(t)$ are in
quantitatively good agreement with numerically exact results when the
driving strength increases from a small value to $A \sim \Xi_0$ or
even to $A=1.5 \Delta$.

From all the above figures, one can see that the CHRW captures
correctly the novel characters of the on-resonance and
near-resonance dynamics. By comparison with the numerical results, the
CHRW treatment gives a significantly better description than
the other treatments.
Physically, the CHRW method takes into account the effects of the bias and the
driving on equal footing. The combined effects result in
the renormalization of the physical quantities in the transformed CHRW
Hamiltonian [see Eq.~(\ref{CHRW})]. For example, in the near-resonance
case of $\omega/\Xi_0=1.2$ ( $\epsilon/\Delta=0.4$ ) with
$A/\Delta=1.3$ of Fig.~\ref{fig2}(d), we obtain $\zeta=0.1855$ and
$\xi=0.6279$ by self-consistently solving Eq.~\ref{xi} and
Eq.~\ref{zeta}. Thus we get the renormalized physical quantities $
\widetilde{A}= 0.5273 \Delta$ and $\widetilde{\Xi}= 1.0085 \Delta$. One
can see from Fig.~\ref{fig2}(d) that the time evolution of the CHRW
method is quantitatively
in good agreement with the numerically exact result, but the Rabi-RWA,
averaged second order VV and RWA-RF results show large deviation from
the numerically exact result. Due to the renormalization, the TLS
$P_{\mathrm{up}}(t)$ of Eq.~(\ref{pt-CHRW}) yields the correct
driven tunneling dynamics. Thus the CHRW method is a simple tractable
method that allows us to study the influence of the bias and CR
terms on the dynamics and the physics in the parameter regime
where the Rabi-RWA and RWA-RF methods fail, especially the moderately strong
driving strength regime with $A\sim \omega \sim \epsilon \sim \Delta$.

\subsection{off-resonance}

\begin{figure}[htbp]
  \includegraphics[width=8cm]{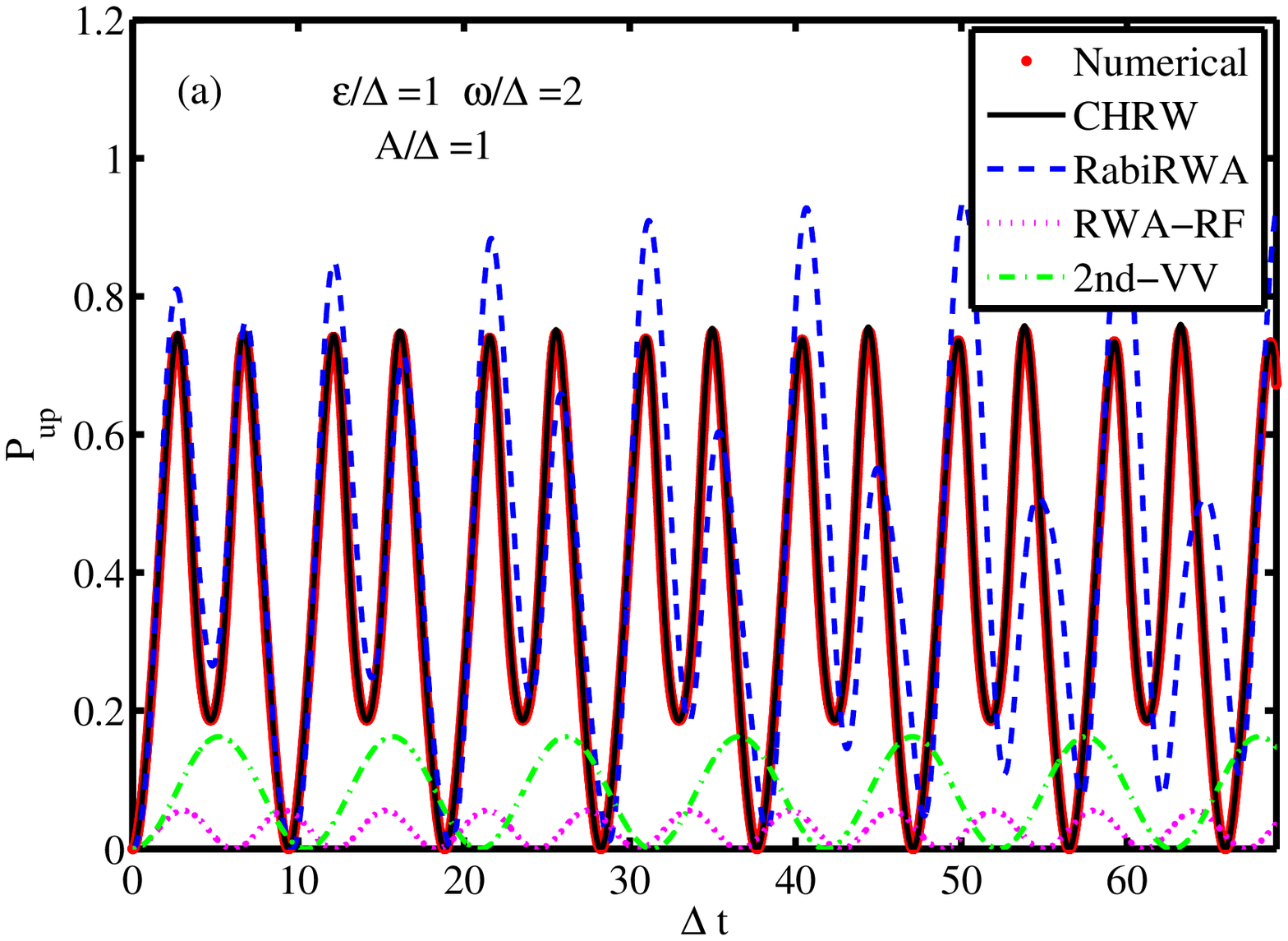}
  \includegraphics[width=8cm]{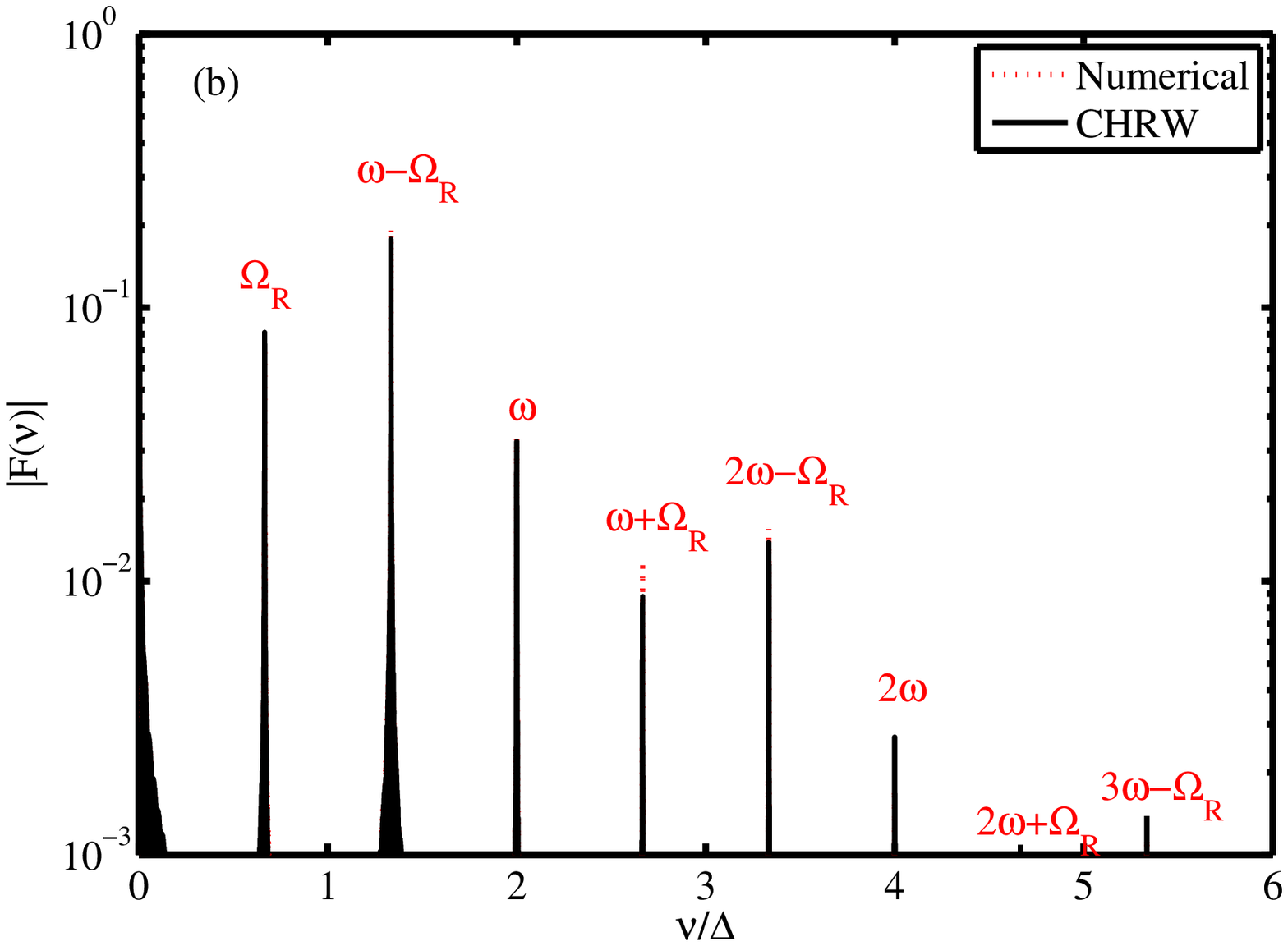}
  \includegraphics[width=8cm]{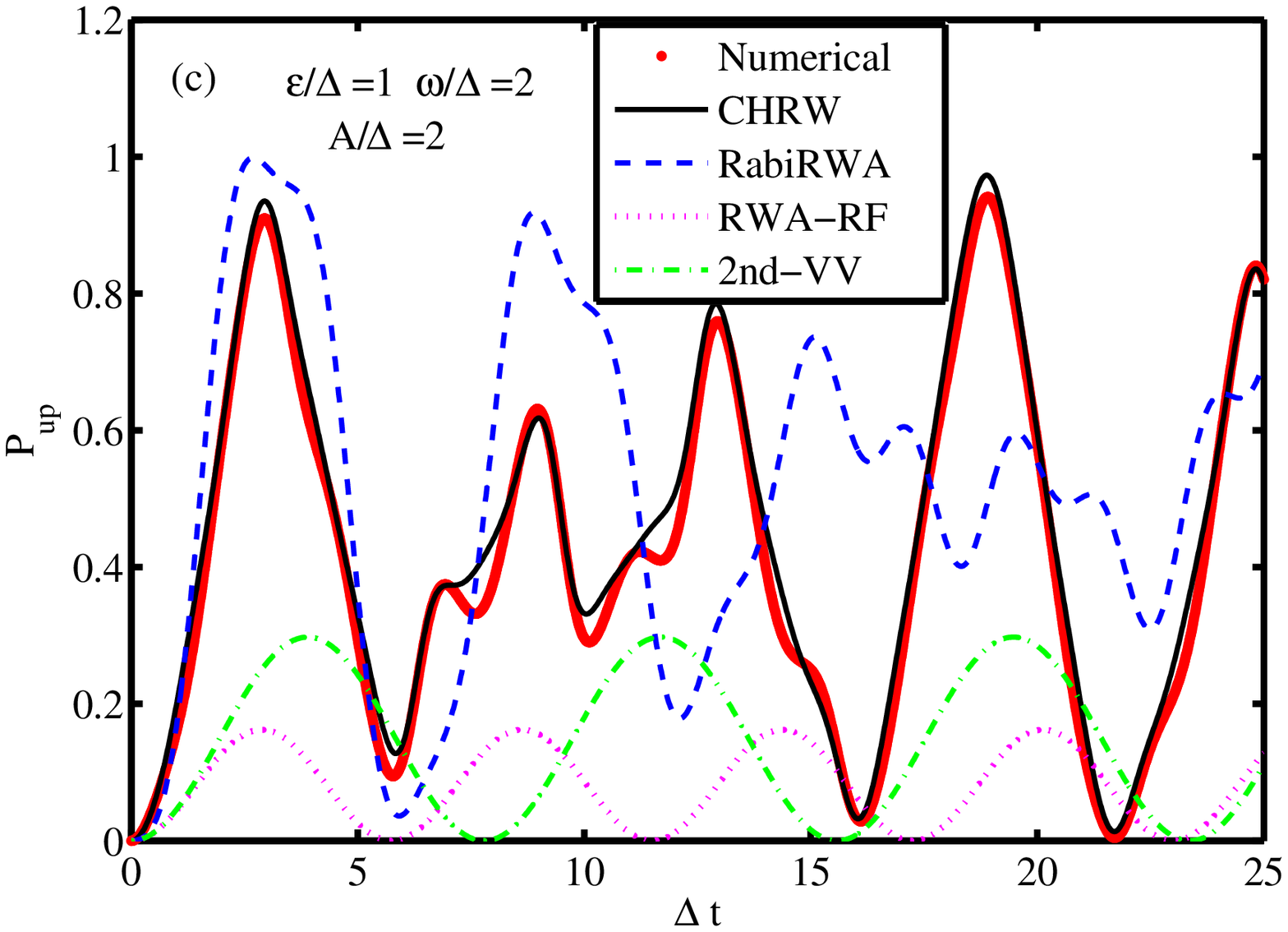}
  \includegraphics[width=8cm]{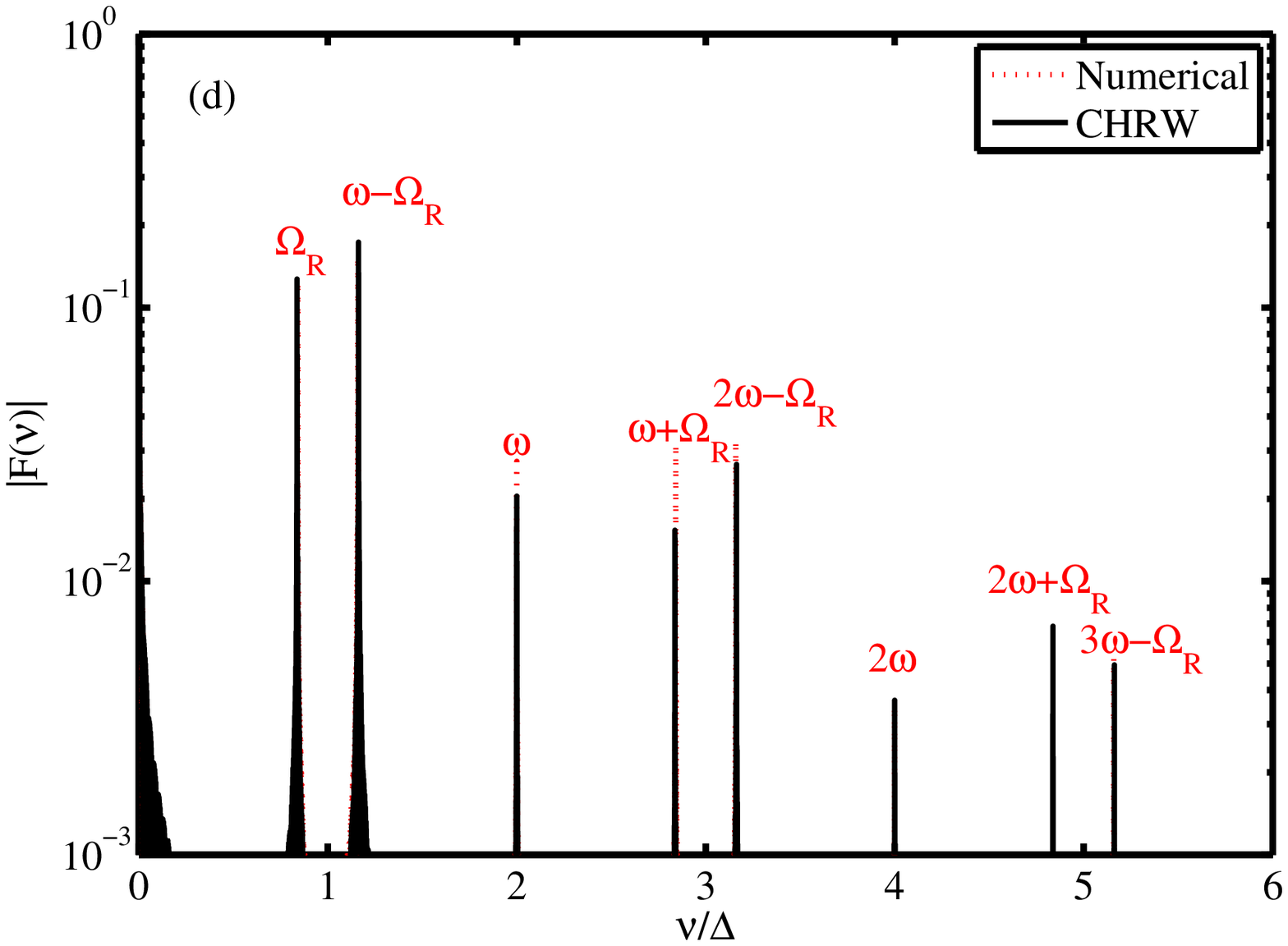}\\
  \caption{(Color online) Time evolutions of $P_{\rm
      up}(t)=\langle\frac{\sigma_z(t)+1}{2}\rangle$ as a function of
    $\Delta t$ for different values of the driving strength (a)
    $A/\Delta= 1$, and (c) $A/\Delta=2$ in the off-resonance case
    ($\omega=2\Delta>\Xi_0$). The corresponding Fourier transform
    $F(\nu)$  of $P_{\rm
      up}(t)$ in (a) and (c) is shown in (b) and (d), respectively.
The bias $\epsilon$ is set to be $\epsilon/\Delta=1$.  }\label{fig3}
\end{figure}

\begin{figure}[htbp]
  \includegraphics[width=8cm]{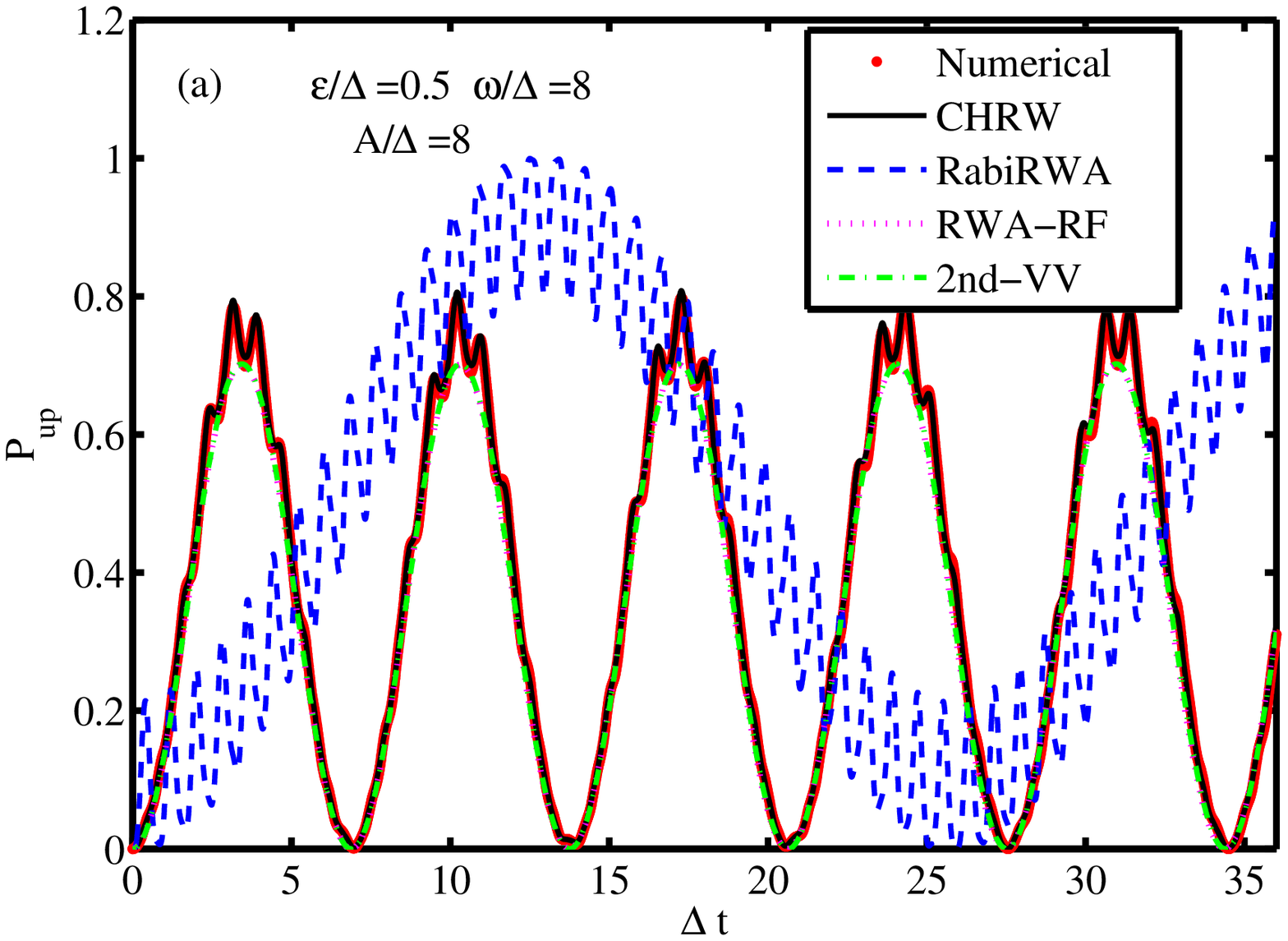}
  \includegraphics[width=8cm]{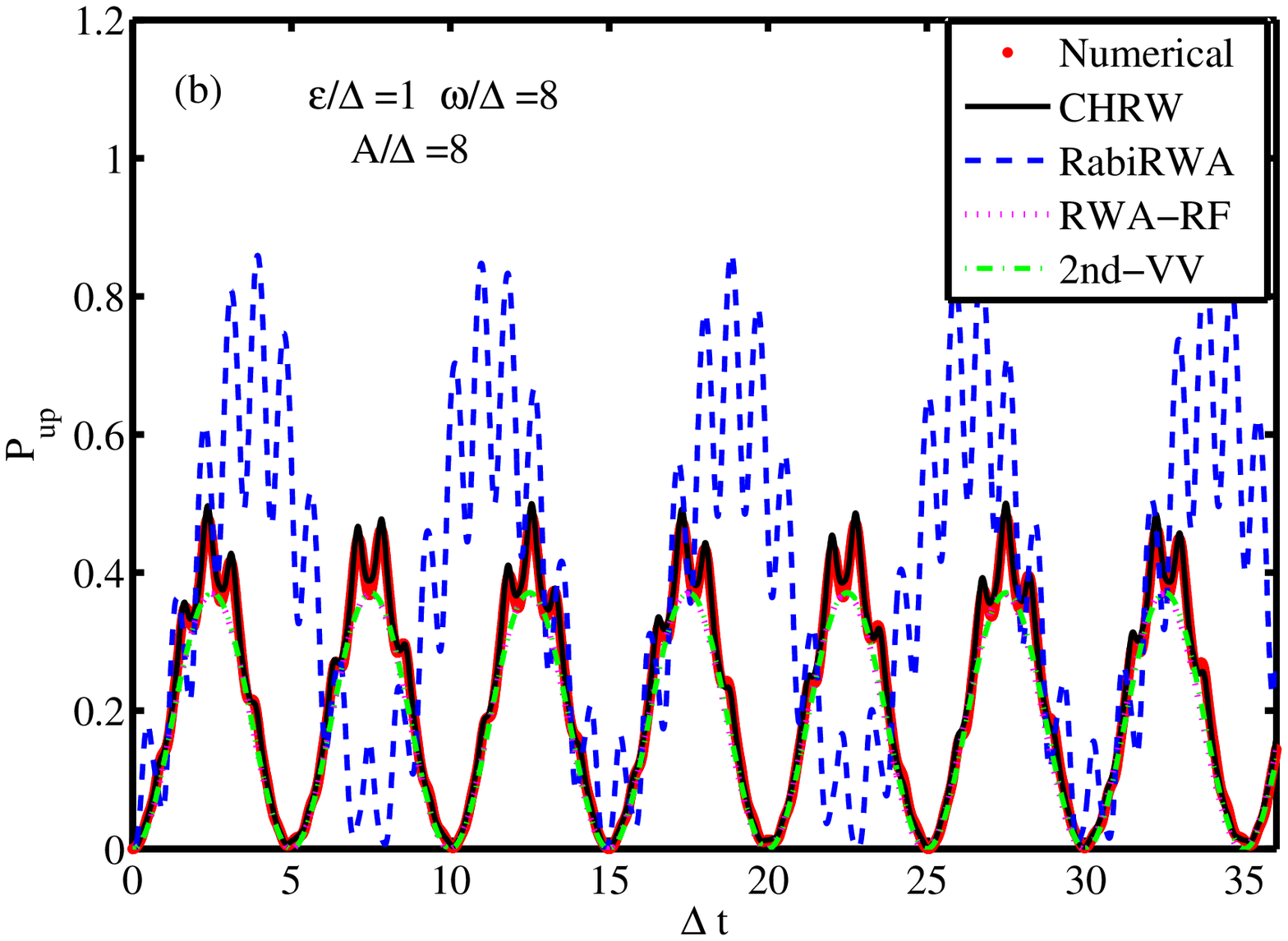}
  \includegraphics[width=8cm]{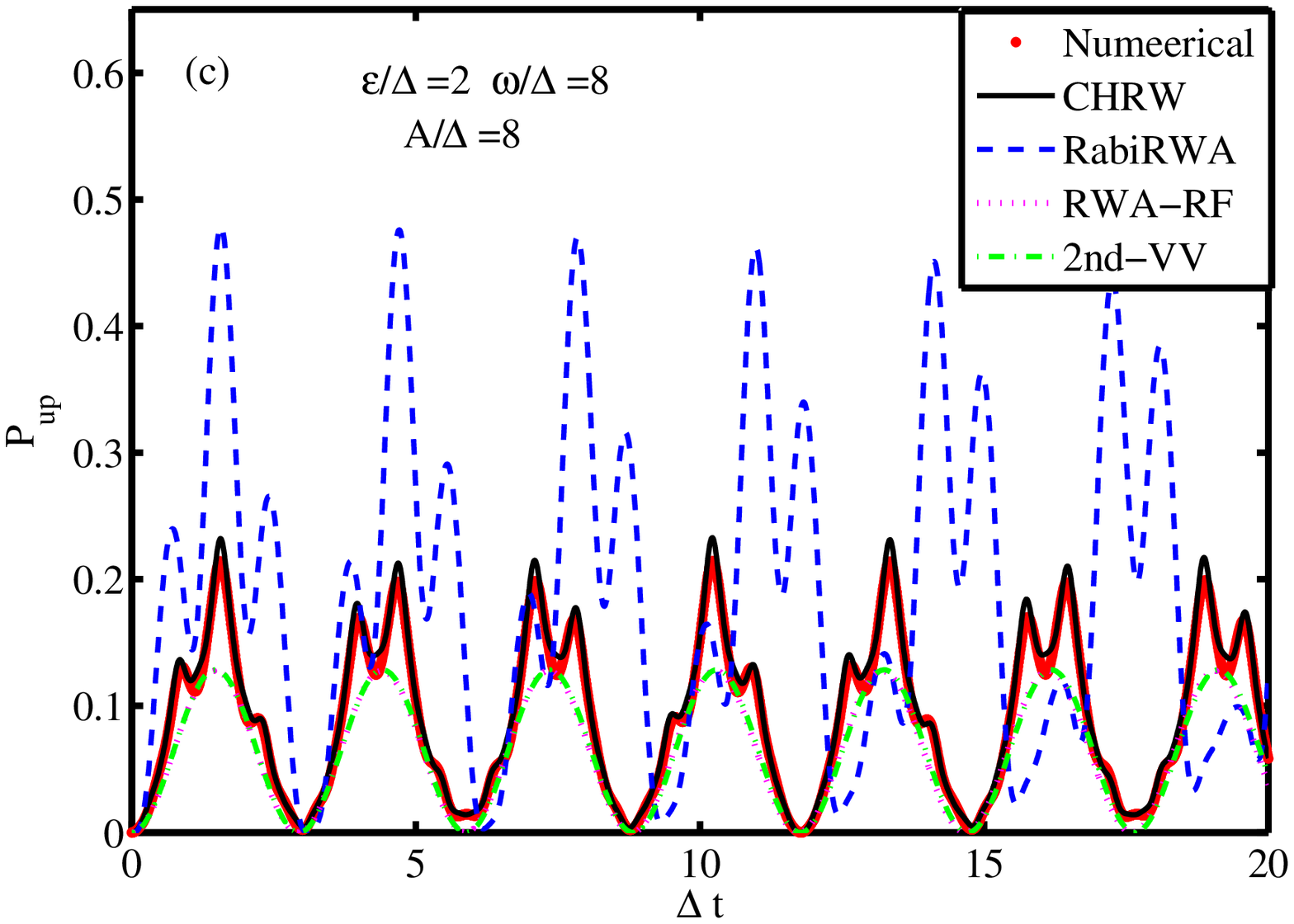}
  \includegraphics[width=8cm]{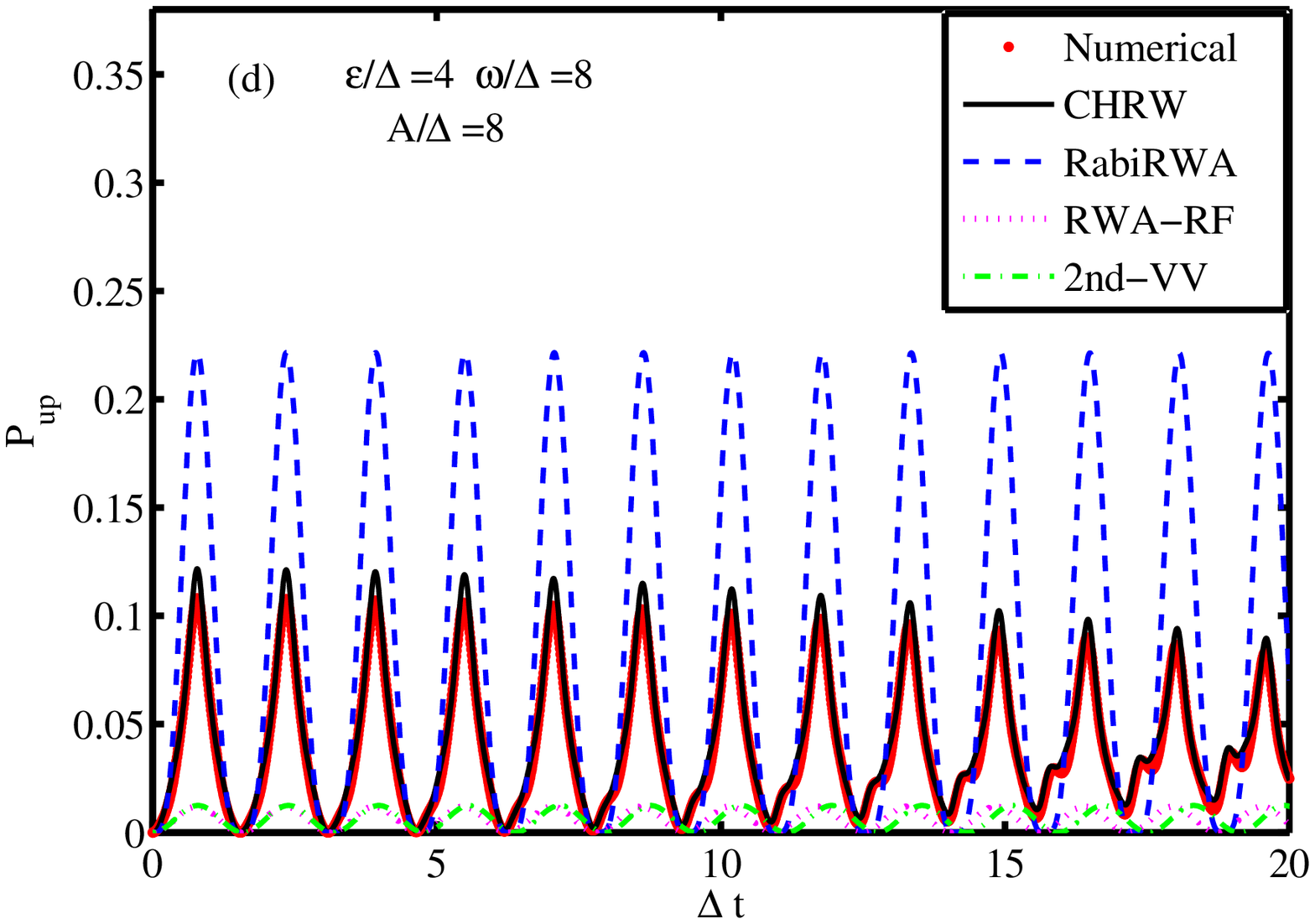}
  \includegraphics[width=8cm]{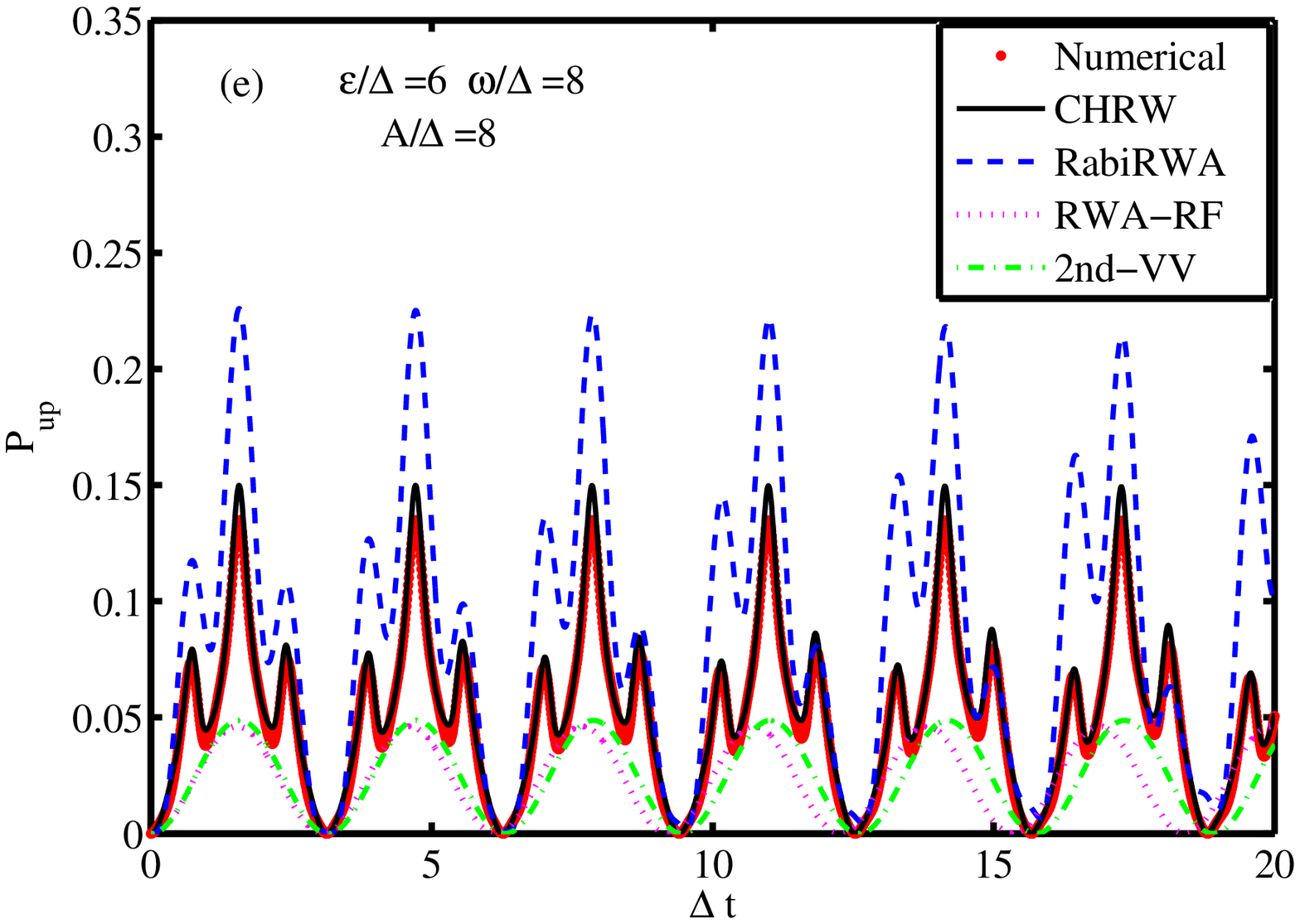}
  \includegraphics[width=8cm]{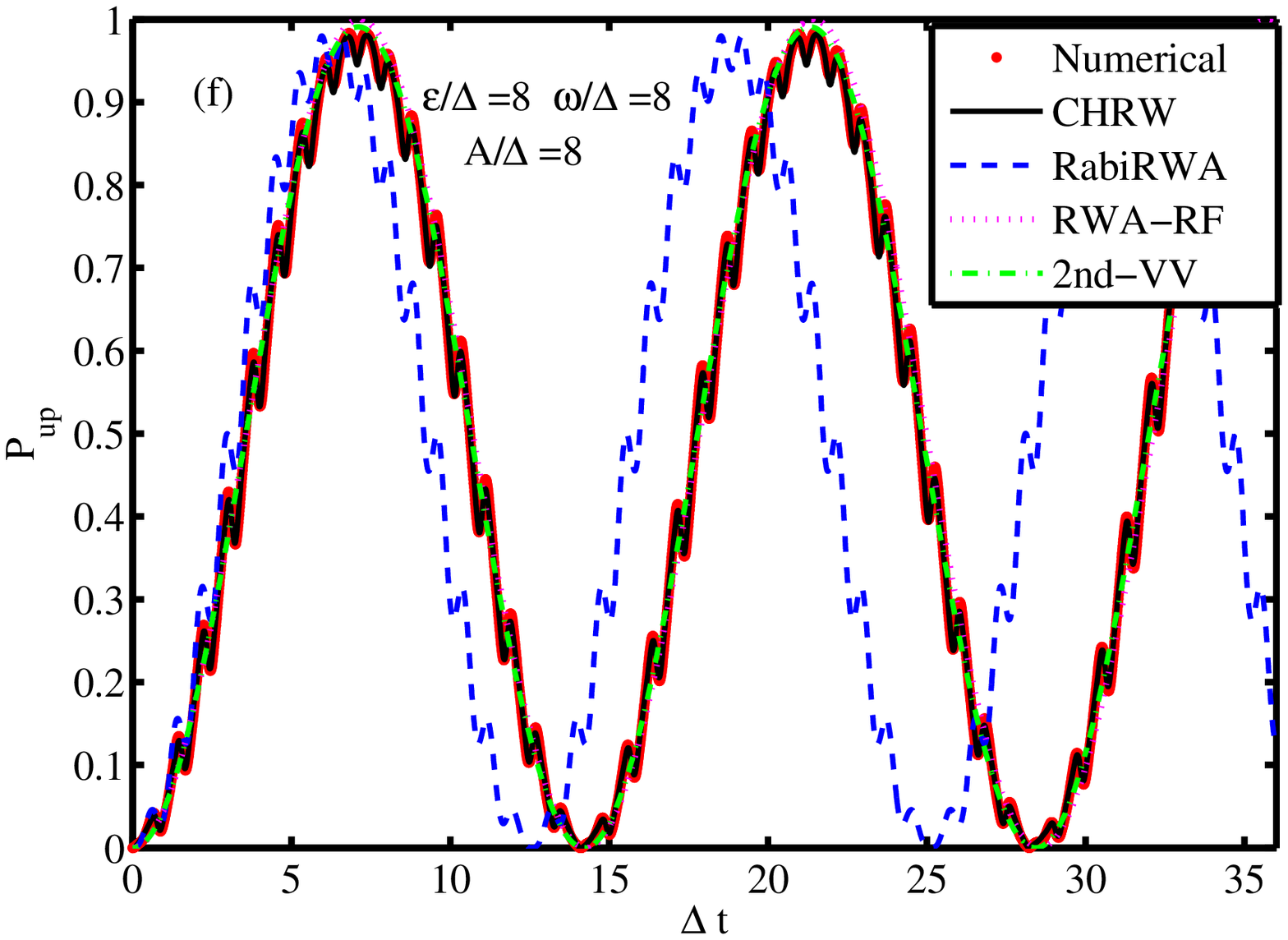}\\
  \caption{(Color online) Time evolutions of $P_{\rm
      up}(t)=\langle\frac{\sigma_z(t)+1}{2}\rangle$ as a function of
    $\Delta t$ for different values of the bias
   (a) $\epsilon /\Delta= 0.5$, (b) $1$, (c) $2$, (d) $4$, (e) $6$,
   and (e) $8$ in
    the off-resonance case ($\omega=8\Delta \neq \Xi_0$).
The driving strength $A$ is set to be
    $A/\omega=1$ ($A/\Delta=8$).  }\label{fig4}
\end{figure}

\begin{figure}[htbp]
  \includegraphics[width=8cm]{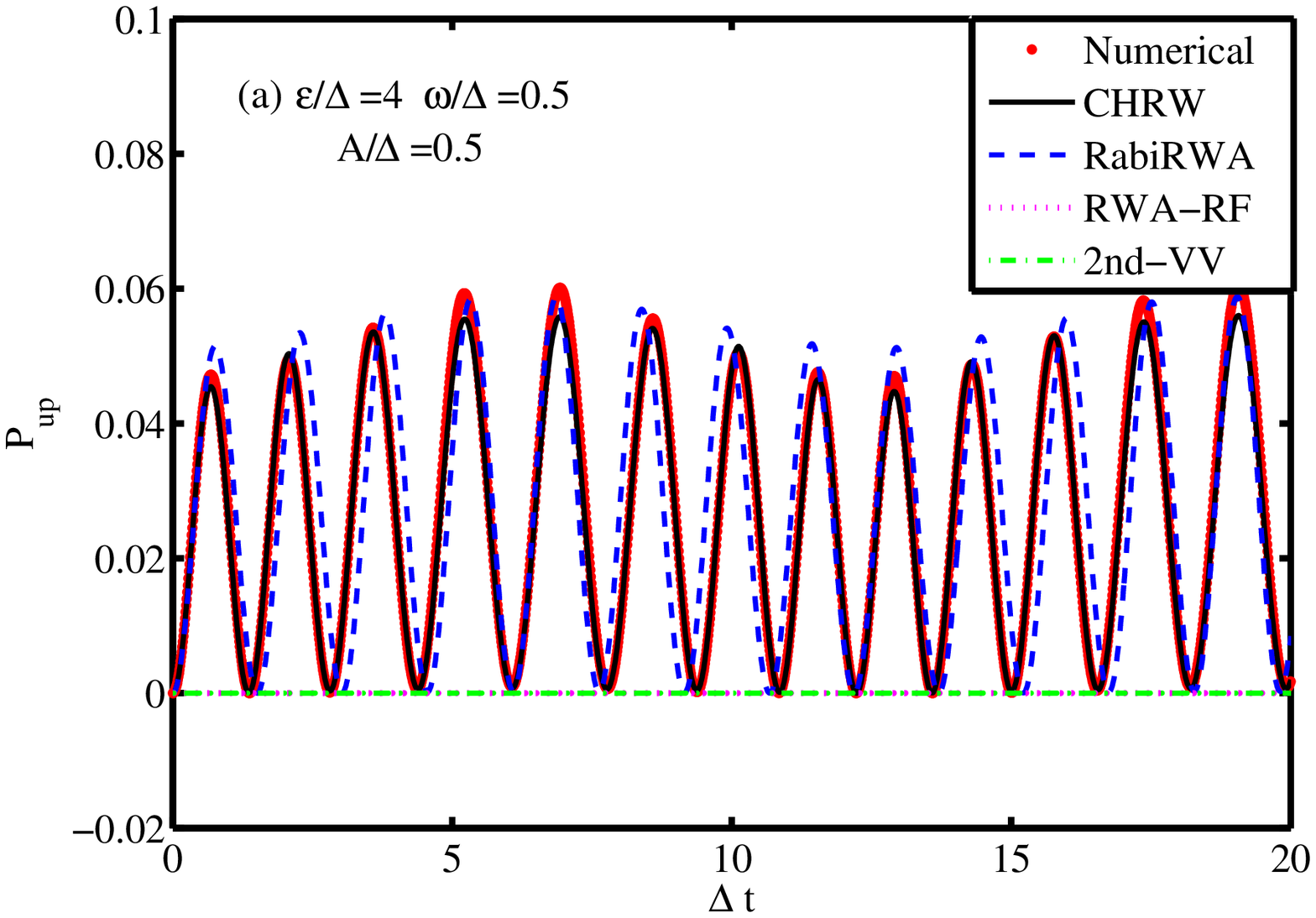}
  \includegraphics[width=8cm]{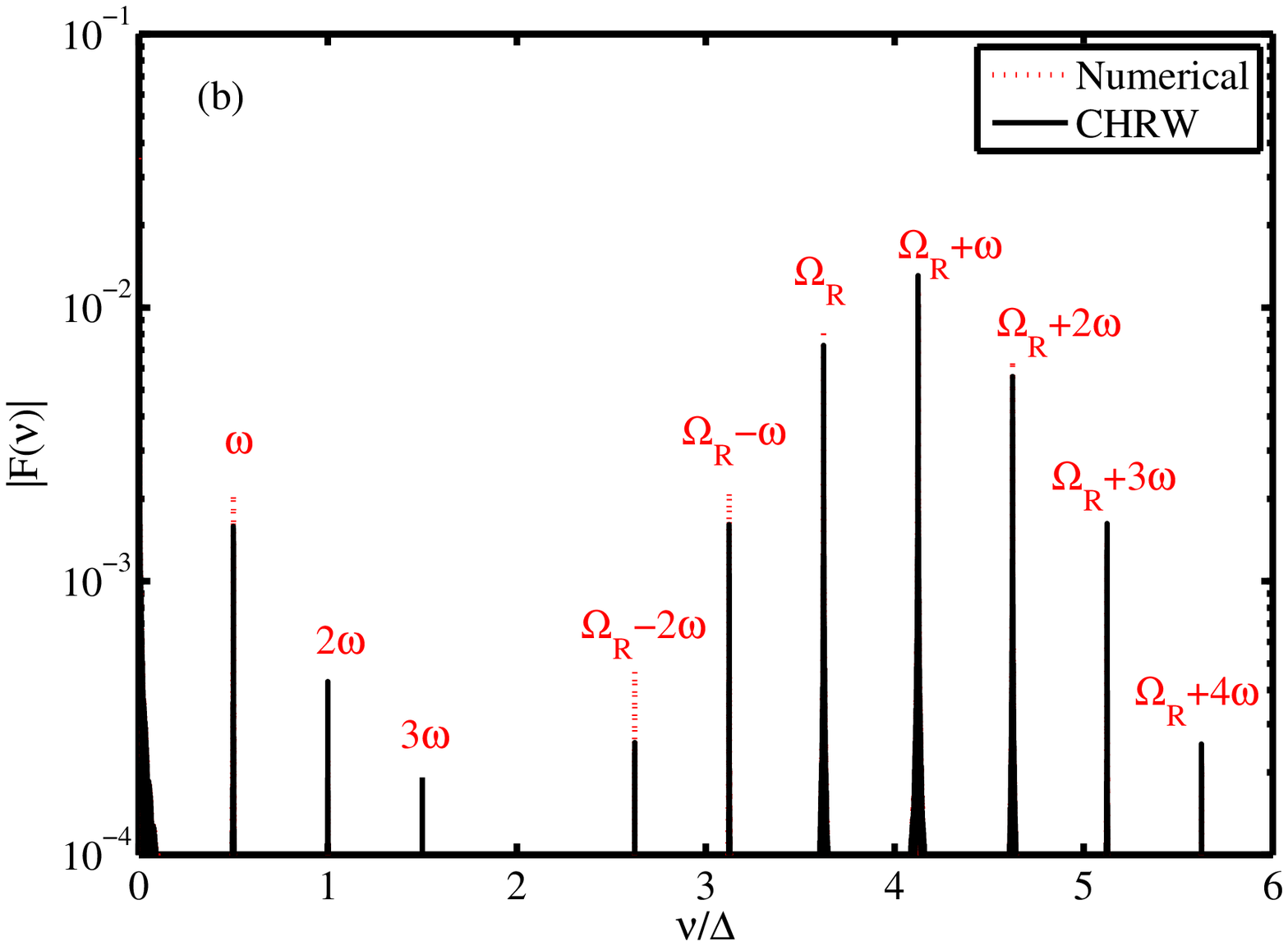}
  \includegraphics[width=8cm]{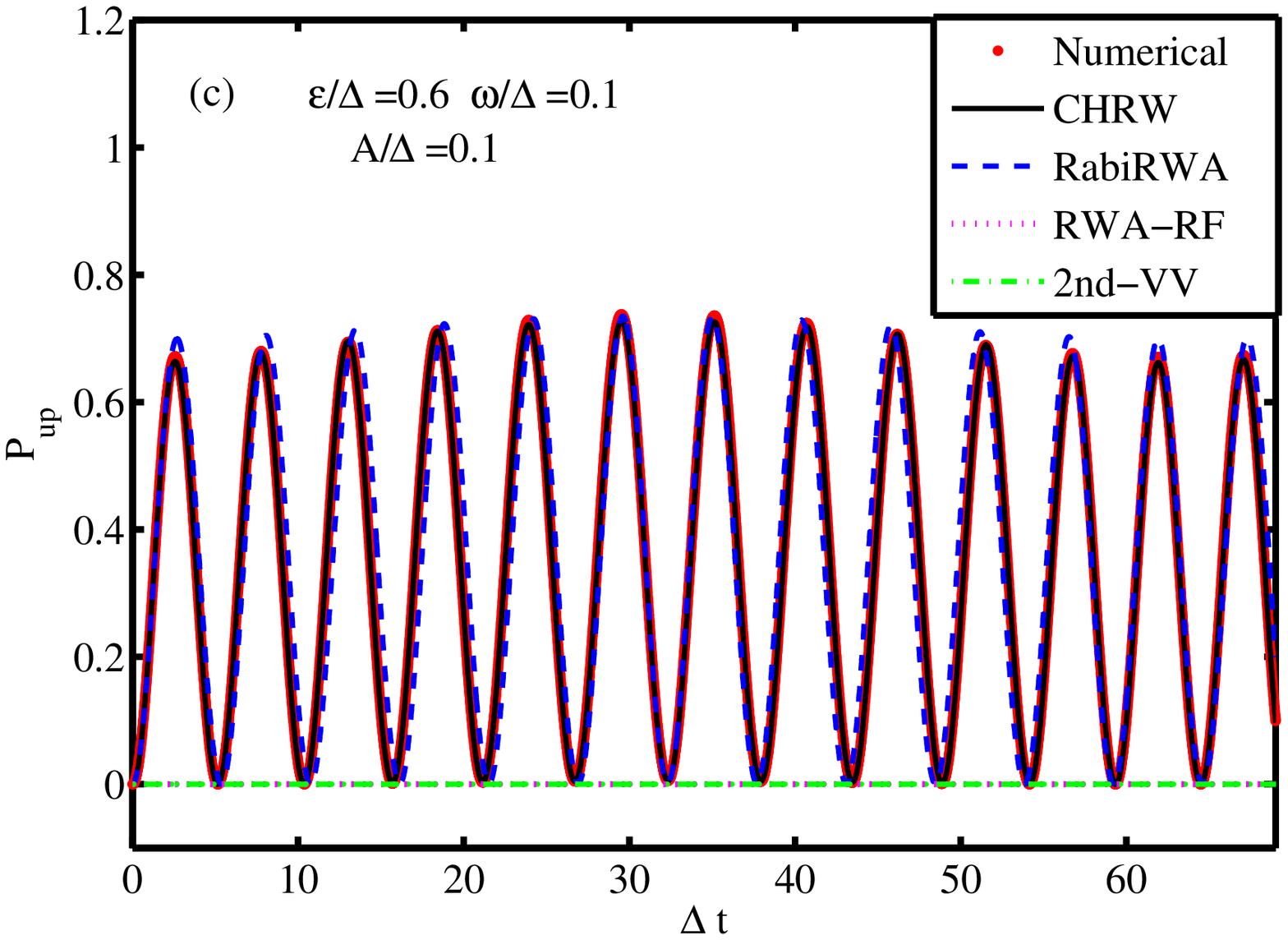}
  \includegraphics[width=8cm]{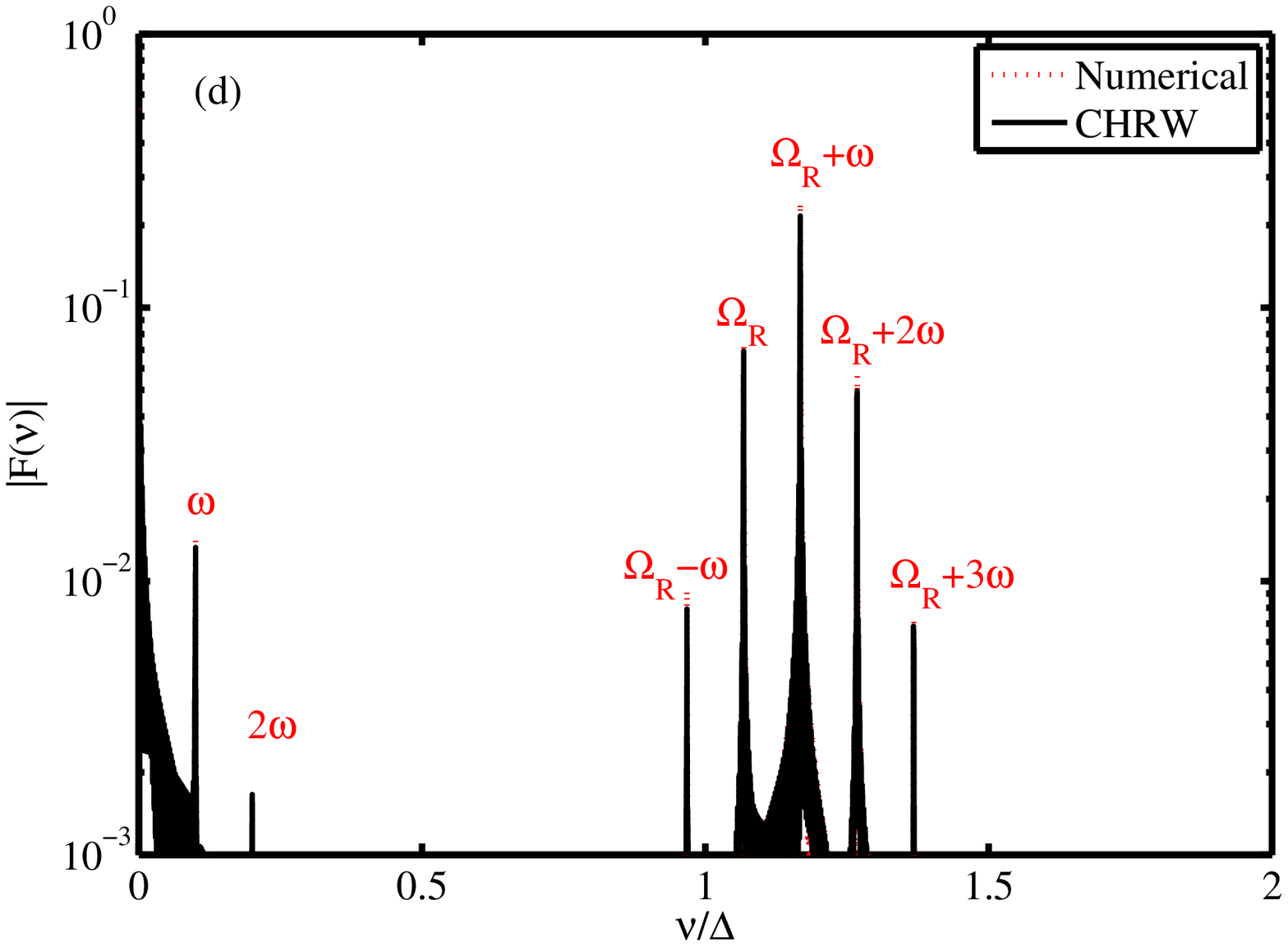}\\
\caption{(Color online) Time evolutions of $P_{\rm
      up}(t)=\langle\frac{\sigma_z(t)+1}{2}\rangle$ as a function of
    $\Delta t$ for different values of the bias (a) $\epsilon/\Delta=4$,
    and (b) $0.6$ in the off-resonance case
    ($\omega<\Xi_0$).  The corresponding Fourier transform  $F(\nu)$ of $P_{\rm
      up}(t)$ in (a) and (c) is shown in (b) and (d), respectively.
The driving strength is set to be $A/\omega=1$.  }\label{fig5}
\end{figure}

Next, we show the time evolutions of
$P_{\mathrm{up}}(t)$ in the off-resonance case of
$\omega=2\Delta>\Xi_0$ with $\epsilon/\Delta=1$
for two moderately strong driving strengths $A/\Delta=1$ and
$A/\Delta=2$ in Fig.~\ref{fig3}(a) and (c), respectively.
At the same time, we also show
the corresponding Fourier transform with a discrete set of frequency
components in Fig.~\ref{fig3}(b) and (d).
For $A/\Delta=1$, the dynamics of the the CHRW method
agrees quite well with the numerically exact ones, which can be
confirmed by the consistence of their frequency components as shown
in Fig.~\ref{fig3}(b). The dynamics of the Rabi-RWA exhibits some
deviation in oscillation amplitude from but
with the main oscillation frequency close to the numerical results in
Fig.~\ref{fig3}(a). On the other hand, the main oscillation
frequencies of the RWA-RF and second-order VV results are
substantially different from that of the exact result.
As the driving strength increases up to
$A/\Delta=2$, the CHRW method still gives a correct dynamics with
only small errors in amplitude [see Fig.~\ref{fig3}(c) and (d)].
In contrast, the other analytical methods
could not give the accurate oscillations and Rabi frequency.

In Fig. \ref{fig4}, we show the effects of the bias on the dynamics of
the driven system for the off-resonance ($\omega=8\Delta \neq \Xi_0$)
and strong driving ($A=\omega=8\Delta$) cases. In comparison with the exact
results, the CHRW method gives not only the main oscillations but also
the time evolution of tiny higher harmonic frequency right for all the
cases of Fig.~\ref{fig4}(a)-(f). In
contrast, the Rabi-RWA approach fails apparently in the strong driving
cases.
The RWA-RF and the 2nd-VV results have the main oscillations close to
those of the
numerical results for almost all the cases, while they can not show
the fine structures in the time evolutions. Furthermore, the
oscillation amplitudes of the RWA-RF and the 2nd-VV results in
Fig.~\ref{fig4}(d) and (e) are considerably smaller than those of the
numerical results.

We show the time evolutions of $P_{\mathrm{up}}(t)$
as well as the corresponding Fourier transform
for the off-resonance case of $\omega<\Xi_0$
in Fig.~\ref{fig5}. In the large
bias case of $\epsilon=4\Delta$ in Fig.~\ref{fig6}(a) with
$A=\omega=0.5 \Delta$, we solve the Rabi frequency,
Eq.~(\ref{eq:Rabi_freq}), and obtain
$\Omega_{\rm R}/\Delta=3.6238 \gg \omega/\Delta=0.5$.
The frequency of amplitude envelope of the driven dynamics in
Fig. \ref{fig5}(a) is $0.5\Delta$.
On the other hand, in the small
bias case of $\epsilon=0.6\Delta$ in Fig.~\ref{fig5}(a) with
$A=\omega=0.1\Delta$, we get the Rabi frequency
$\Omega_{\rm R}/\Delta=1.0677 \gg \omega/\Delta=0.1$.  By the Fourier
transform of the time evolutions, we show in
Figs.~\ref{fig5}(b) and (d) the discrete
frequencies are $n\omega$ and $\Omega_{\rm R} \pm n\omega$
($n=0,1,2,...$). The dominate
oscillation frequency is neither $\omega$ nor $\Omega_{\rm R}$. It is
$\Omega_{\rm R}+\omega$ with the largest weight for the parameters
in Fig.~\ref{fig5}.
Besides, the Rabi-RWA results agree roughly with the exact
results, but the RWA-RF and 2nd-VV results show $P_{\rm up}(t)=0$ without
any oscillation in contrast with the numerical results. In other
words, in this parameter regime of $A\omega/\Delta^2<1$, the RWA-RF and
the 2nd-VV methods are invalid.

\section{Generalized Rabi frequency} \label{sec.Rabi}

We discuss and calculate
the generalized Rabi frequency and the Bloch-Siegert shift
and compare our calculated
values with the data shown in the experiment of flux qubit
\cite{Yoshihara}. First, we derive, for simplicity, a formula of
the Rabi frequency to second order in $A$
based on the general Rabi frequency, Eq.~(\ref{eq:Rabi_freq}).
Then we calculate and discuss the bias-modulated Bloch-Siegert shift.
Finally, we illustrate the valid
parameter regime of our CHRW method.

\subsection{ Frequency shift: Bloch-Siegert shift }

In the following, we shall address the question whether the bias leads
to a profound change of the Rabi frequency and the shift of the resonance
frequency. To this end, in this subsection we calculate the generalized
Rabi frequency and bias-modulated Bloch-Siegert shift analytically and
numerically.
From Eqs.~(\ref{eq:Rabi_freq}), (\ref{eq:ren_detuning}),
(\ref{A_ren}), (\ref{eq:Xi_ren}), (\ref{Delta_ren}) and (\ref{epsilon_ren}),
the general renormalized Rabi frequency  that takes into account the
effects of CR couplings and the bias
 can be written as
\begin{eqnarray}\label{Rabifreq-CHRW}
  \Omega_{\rm R}^2 &=& \left[\omega - \widetilde{\Xi} \right]^2 + \left[2 \frac{\Delta \xi -\epsilon \zeta}{X}J_{1}\left(\frac{A}{\omega} X\right)\right]^2 \nonumber \\
  &=& \left\{\omega-\sqrt{\Xi_0^2-\left[1-J^2_{0}\left(\frac{A}{\omega} X\right) \right]\left(\frac{\Delta \xi -\epsilon \zeta}{X}\right)^2}\right\}^2 + \left[2 \frac{\Delta \xi -\epsilon \zeta}{X}J_{1}\left(\frac{A}{\omega} X\right)\right]^2.
\end{eqnarray}
For a finite bias, we will derive an analytical expression up to second
order in the driving strength $A$. The Bloch-Siegert shift is a well-known
correction to the RWA, and accounts for the CR field to leading
order.
Expanding $\xi$ and $\zeta$ up to lowest order in $A$, we obtain from Eqs.(\ref{xi}) and (\ref{zeta})
\begin{eqnarray}
 \Delta\xi-\epsilon \zeta &=& \frac{\omega \Delta}{\omega+\Xi_0} .
\end{eqnarray}
Thus the modulated effective Rabi frequency
to second order in $A$ has the form
\begin{eqnarray}\label{Rabifreq}
  \Omega_{\rm R-2nd}^2 &=& \left[\omega-\Xi_0\right]^2+\frac{A^2\Delta^2}{2\Xi_0(\omega+\Xi_0)},
\end{eqnarray}
whose expression in the limit of vanishing bias is the same as those
given in  Refs.~\cite{GH} and \cite{BS-shift}. Moreover,
Eq.~(\ref{Rabifreq}) can be used to analytically calculate the
Bloch-Siegert shift of the resonance frequency $\Omega_{\mathrm{res}}$
which is defined as the frequency at which the transition probability
average is a maximum \cite{BS-shift}. This occurs when $\partial
\Omega_{\rm R}^2/ \partial \Delta =0$\cite{BS-shift}. Thus, we obtain
the Bloch-Siegert shift $\delta\omega_{\mathrm {BS}}$ as
\begin{equation}\label{BS}
    \delta\omega_{\mathrm {BS}}=\Omega_{\mathrm{res}}-\Xi_0= \frac{A^2}{4\Xi_0}\left[1-\frac{3}{4}\left(\frac{\Delta}{~\Xi_0}\right)^2\right].
\end{equation}
For $\epsilon=0$, Eq.~(\ref{BS})
reproduces the result of $\frac{1}{16}\frac{A^2}{\Delta}$
given in Ref.~\cite{BS-shift}. In the unbiased case, the Rabi
frequency $\Omega_{\rm R}$ up to fourth
order in $A$ has been given in our previous work
\cite{ZG2012}. Moreover, we confirm
that the Bloch-Siegert shift for the unbiased case given
by our method is in a good agreement with that obtained by the Floquet
approach in the entire driving-strength regime \cite{Yan15}. These
results strongly prove that the CHRW method has properly taken into
account the effects of CR terms and the bias.

\begin{figure}[htbp]
  \includegraphics[width=8cm]{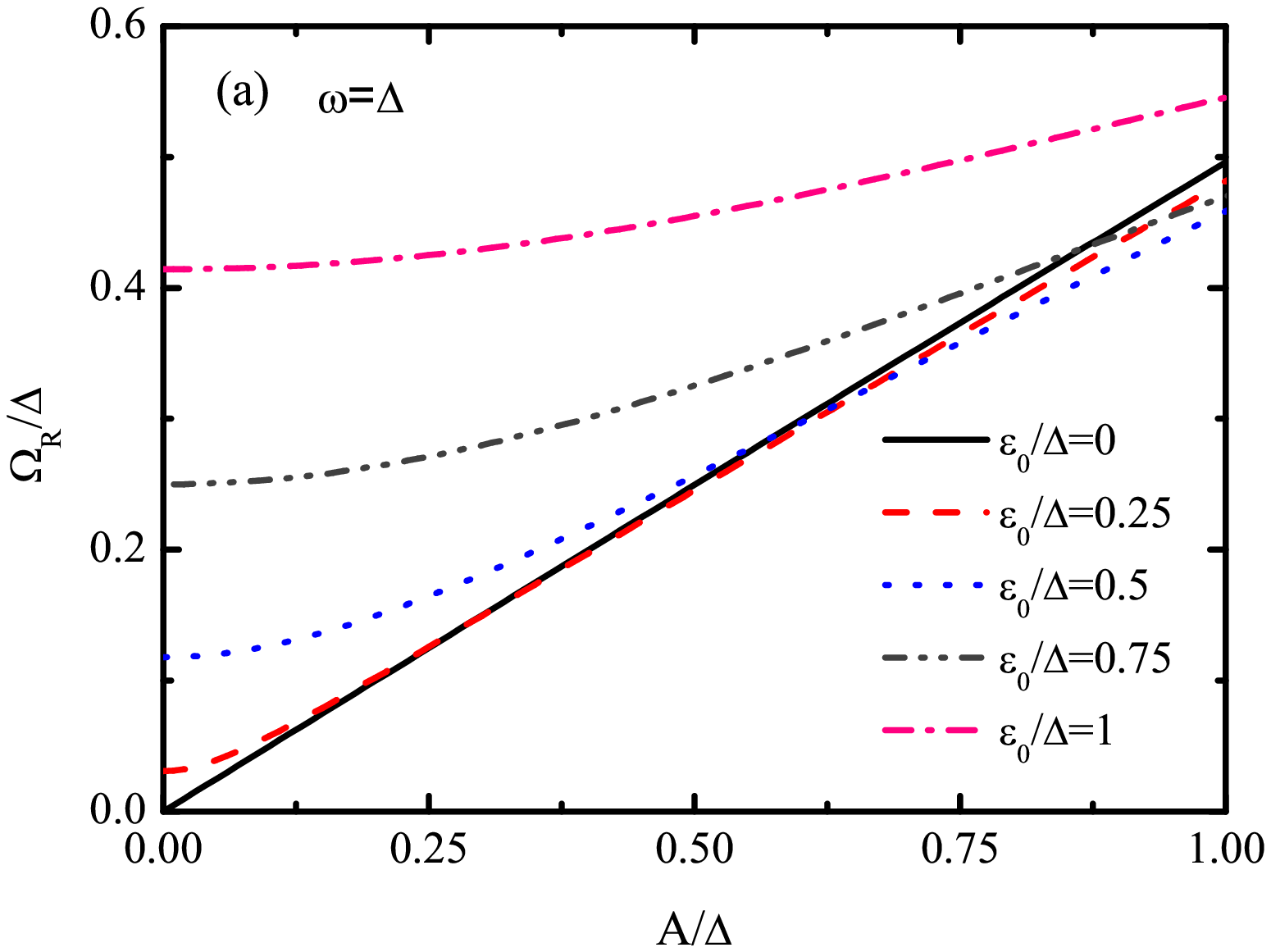}
  \includegraphics[width=8cm]{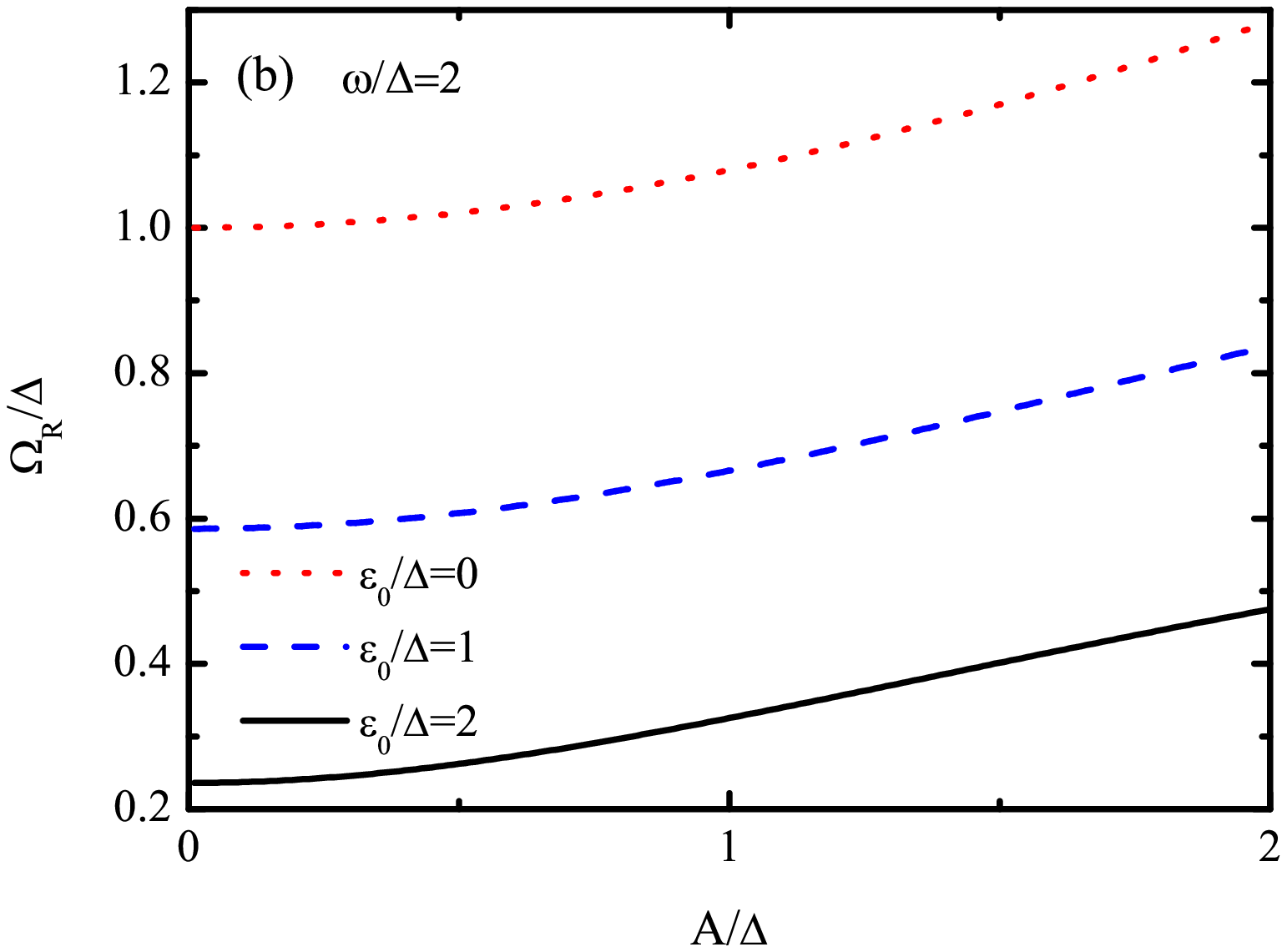}\\
  \includegraphics[width=8cm]{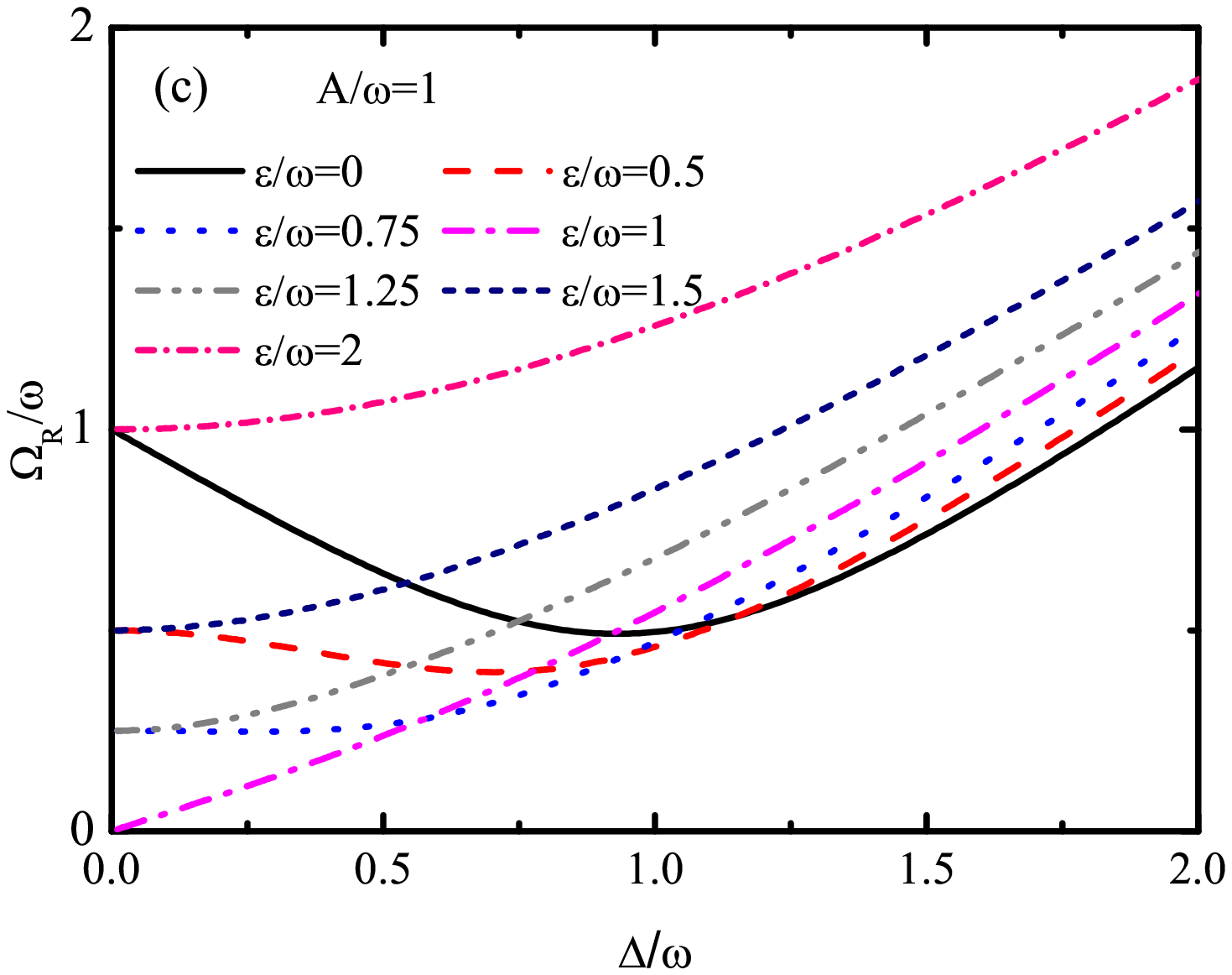}
  \includegraphics[width=8cm]{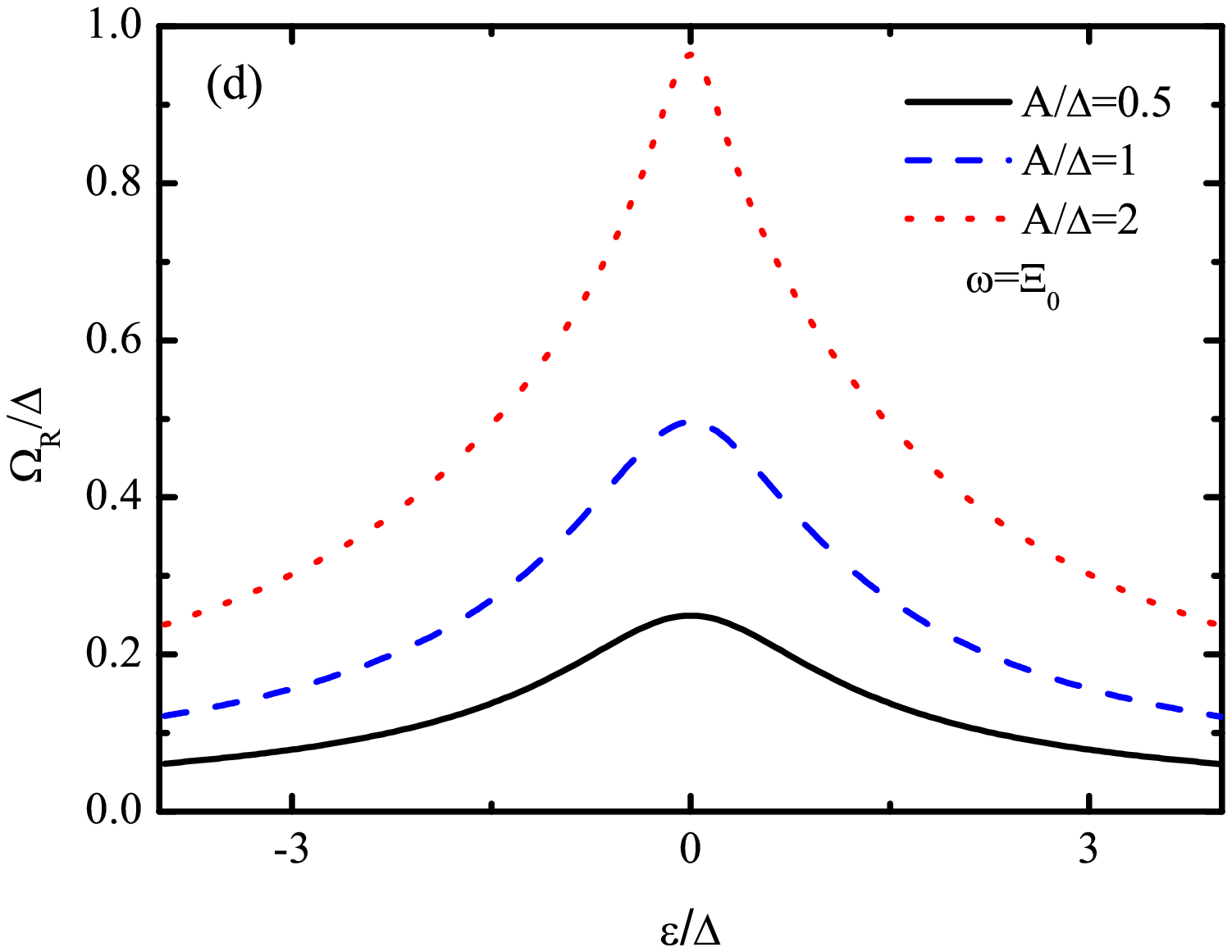}\\
  \caption{(Color online) (a) Generalized Rabi frequency $\Omega_{\rm R}$ as
    a function
    of the driving strength $A$ with $\omega=\Delta$ for  several
    different values of bias $\epsilon/\Delta=0$, $0.25$, $0.5$,
    $0.75$, and $1$. (b) Generalized Rabi frequency $\Omega_{\rm R}$
    as a function of
    $A$ with $\omega=2\Delta$ for different values of bias
    $\epsilon/\Delta=0$, $1$, and $2$. (c) Generalized Rabi frequency
    $\Omega_{\rm R}$ as a function of $\Delta$ with $A/\omega=1$ for several
    different values of bias $\epsilon/\omega=0$, $0.5$, $0.75$, $1$, $1.25$,
    $1.5$ and $2$. (d) Generalized Rabi frequency $\Omega_{\rm R}$ as a
    function of the
    bias $\epsilon$ with $\omega=\Xi_0$ for several different values
    of driving strength $A/\Delta=0.5$, $1$, and $2$. }\label{fig6}
\end{figure}

\subsection{ Effects of the bias on the Rabi frequency }

We discuss here the dependence of the generalize Rabi frequency $\Omega_{\rm R}$ on
the parameters of $A$, $\Delta$, and $\epsilon$. Figure \ref{fig6}(a)
shows $\Omega_{\rm R}$ as a function of the driving strength $A$ for various
bias values and a fixed driving frequency $\omega=\Delta$.
One can see that for $\epsilon=0$, $\Omega_{\rm R}$ is linearly proportional
to $A$ in the small driving strength regime where  $A/\Delta \leq 1$.
When $A/\Delta <0.5$,
the values of $\Omega_{\rm R}$ with a finite bias $\epsilon >0 $ are larger than
those with $\epsilon=0$, and increases with the increase of the
bias. When $A/\Delta >0.5$, however, the crossover of the curves for
different values of the bias $\epsilon$ appears. In this regime, the
increase of the bias does not
always favor the increase of $\Omega_{\rm R}$. This comes from the competition
between different controlled parameters ($\epsilon/\Delta, ~\omega/\Delta$, and $A/\Delta$).
It indicates that the relation of $\Omega_{\rm R}$ versus $A$ is beyond a linear dependent behavior when all
the energy scales
are nearly in the same order. Figure \ref{fig6}(b) displays the Rabi
frequency $\Omega_{\rm R}$ as a function of the driving strength $A$ for
three values of bias $\epsilon$ and a fixed
driving frequency $\omega=2\Delta$. Obviously, the increase of the
bias $\epsilon$ decreases the Rabi frequency $\Omega_{\rm R}$ in this
parameter regime.

In Fig. \ref{fig6}(c), we show the dependence of the Rabi frequency
$\Omega_{\rm R}$ on $\Delta$ for different values of the bias $\epsilon$
with $A/\omega=1$. For
$\epsilon=0$, the position of the scaled tunneling $\Delta/\omega$
corresponding to the minimum value of $\Omega_{\rm R}$ is not located at $\Delta/\omega=1$
but at $\Delta/\omega=0.93$. This indicates
that the CR terms lead to the explicit deviation from the RWA
result of $\Delta/\omega=1$ in the intermediate driving strength
regime. For finite bias case ($0 < \epsilon \leq 0.5 \omega$), with the
increase of $\Delta/\omega$, $\Omega_{\rm R}$ falls first and then rises.
For $\epsilon/\omega = 0.75$, $\Omega_{\rm R}$ is insensitive to the change of
the scaled tunneling when $\Delta/\omega \leq 0.5$ in comparison with its fast
increase when $\Delta/\omega \geq 1$. For $\epsilon/\omega \geq 1$,
$\Omega_{\rm R}$ generally increases with the increase of the scaled tunneling.
We notice
that near $\Delta/\omega \sim 1$, $\Omega_{\rm R}$ is almost the same for
$\epsilon/\omega <1$.
Moreover, when $\Delta/\omega >1.25$,
$\Omega_{\rm R}$ increases with the increase of the bias in contrast to its
non-monotonic dependence on bias when $\Delta/\omega <0.5$. This indicates
that the competition between the quantum tunneling $\Delta/\omega$ and
the driving $A/\omega$ leads to the
complicated dependence of $\Omega_{\rm R}$ on the bias $\epsilon$ in the intermediate
driving-strength regime.

In Fig. \ref{fig6}(d), we show the Rabi frequency $\Omega_{\rm R}$ as a
function of the bias $\epsilon$ at $\omega=\Xi_0$ for different
driving strengths. $\Omega_{\rm R}$ displays a parity symmetry with respect
to the bias $\epsilon$, i.e., $\Omega_{\rm R} (-\epsilon)=\Omega_{\rm R}
(\epsilon)$, which can be seen, for example, by Eq.~(37) valid to
second order in $A$.  When the bias $\epsilon$ is fixed, the larger
the driving strength $A$, the larger the Rabi frequency
$\Omega_{\rm R}$. When $A$ is set to a fixed value, $\Omega_{\rm R}$ decreases
with increasing the absolute value of $\epsilon$, and $\Omega_{\rm R}$
reduces more drastically for $A/\Delta=2$ than $A/\Delta=0.5$.

\begin{figure}[htbp]
  \includegraphics[width=8cm]{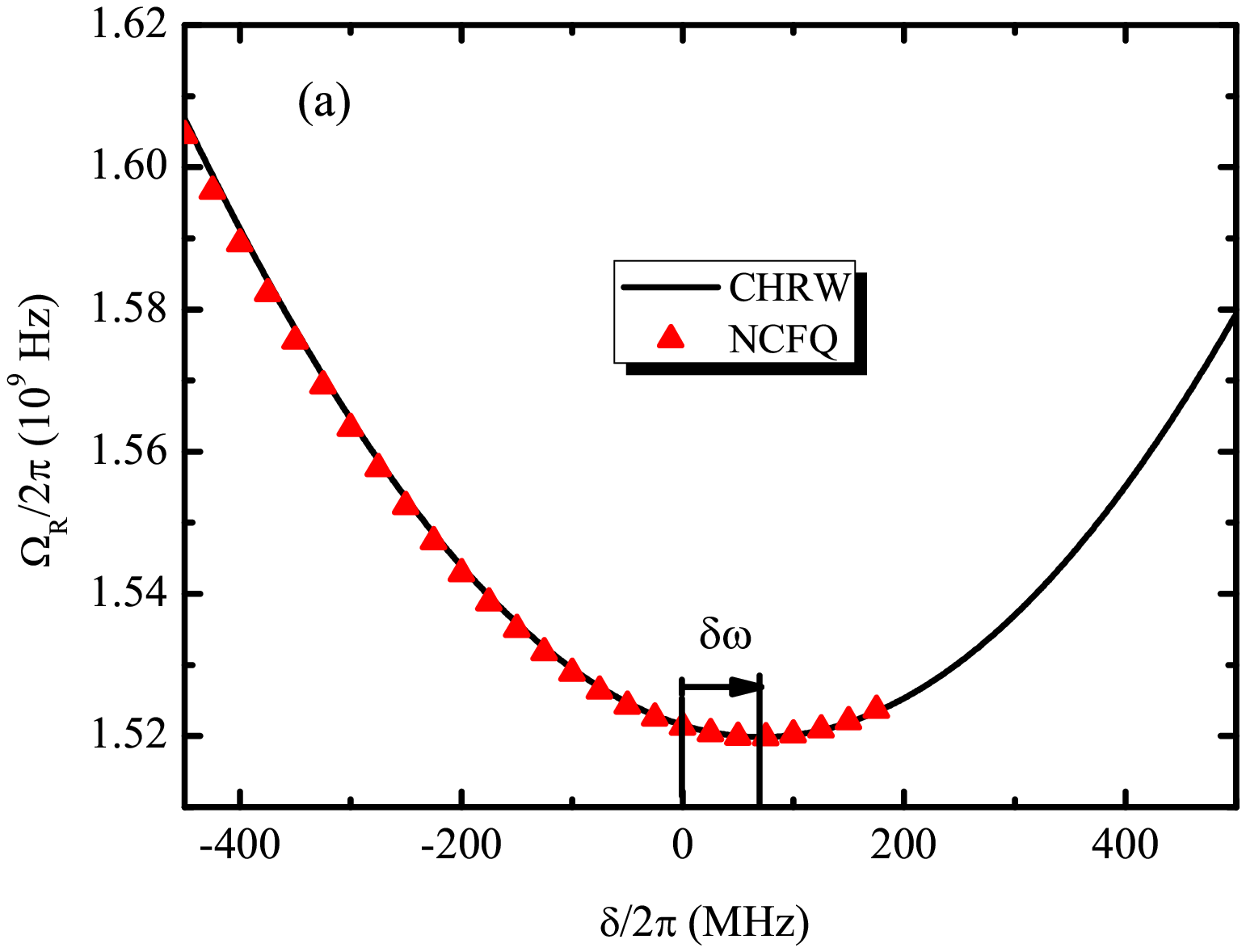}
  \includegraphics[width=8cm]{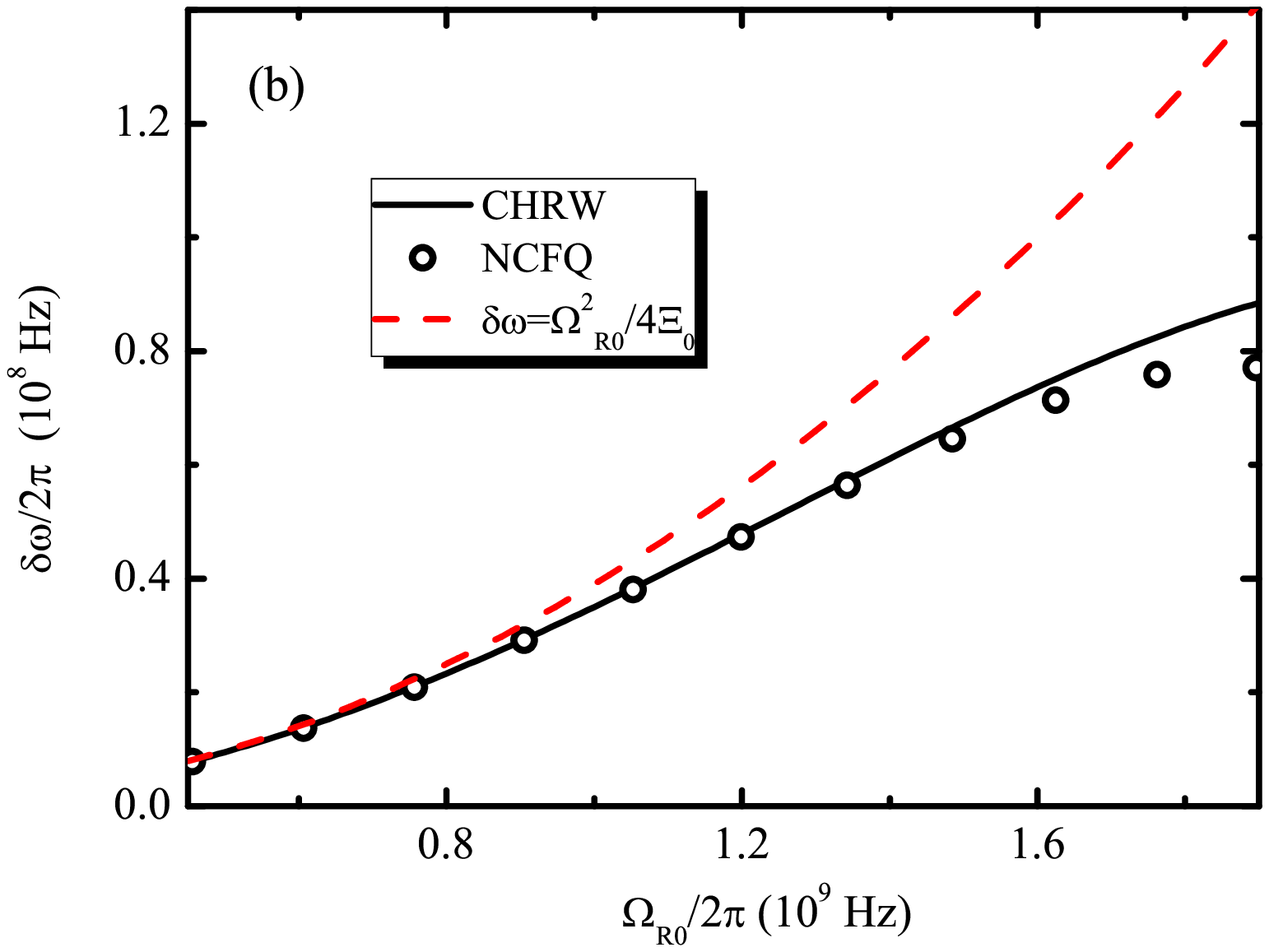} \\
  \includegraphics[width=8cm]{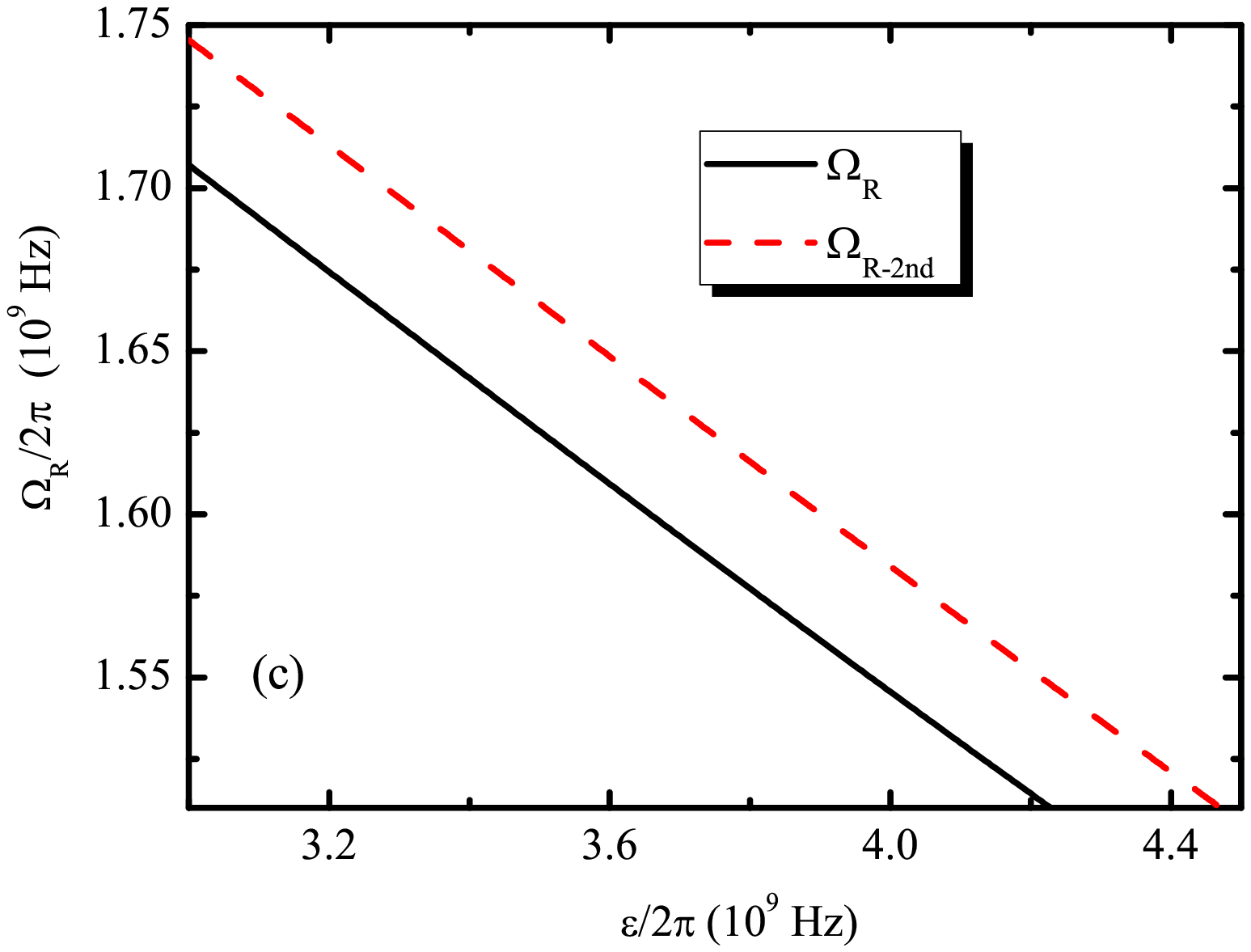}
  \includegraphics[width=4.8cm]{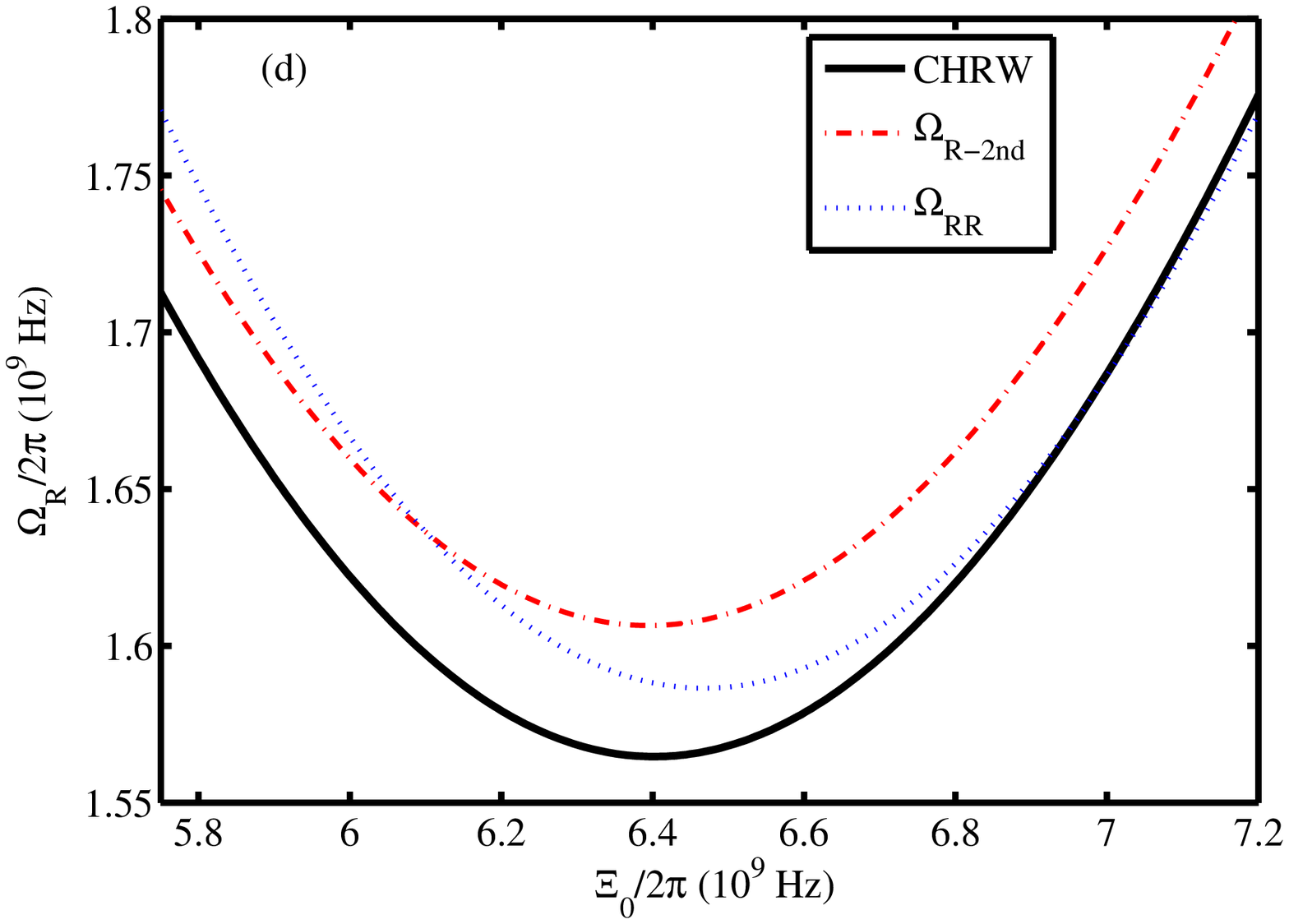} \\
  \caption{(Color online) (a) Generalized Rabi frequency $\Omega_{\rm
      R}$ as a function of the
    detuning $\delta=\Xi_0-\omega$ obtained by the CHRW method (black
    solid line) with $A/2\pi=4.100\rm{GHz}$,
    $\epsilon/2\pi=4.154\rm{GHz}$
    and $\Delta/2\pi=4.869\rm{GHz}$ ($\Xi_0=6.4\rm{GHz}$).
The data for numerically calculation of the flux qubit (NCFQ) in
    Ref.~{\cite{Yoshihara}} are represented by red triangles.
(b) Numerically calculated frequency shift $\delta \omega$
    by the CHRW method (black solid line),
    the second-order Bloch-Sigert shift (red dashed line) and
    the NCFQ results in Ref.~{\cite{Yoshihara}} (open circles). (c)
    Generalized Rabi frequency
    $\Omega_{\rm R}$ as a function of $\epsilon$ given by
    Eq.~(\ref{Rabifreq}) (red dashed line) and
    Eq.~(\ref{Rabifreq-CHRW}) (black solid line). (d) Generalized Rabi
    frequency $\Omega_{\rm R}$ as a function of
    $\Xi_0$ with $\omega/2\pi=6.1 {\rm GHz}$ calculated by
    Eq.~(\ref{Rabifreq}) (red dash-dotted line) and
    Eq.~(\ref{Rabifreq-CHRW}) (black solid line).
     The blue dotted line
    denotes the Rabi frequency $\Omega_{\rm RR}$ obtained by the
    Rabi-RWA method.}\label{fig7}
\end{figure}

Present calculated results may
be examined and compared to available experimental measurements of
superconducting flux qubits.
In order to show the effects of the bias on the generalized Rabi
frequency in the flux qubit, we use the parameters of the flux qubit
in Ref.~{\cite{Yoshihara}}, $\Delta/2\pi=4.869 \rm{GHz}$ and
$\epsilon/2\pi=4.154\rm{GHz}$ ($\Xi_0/2\pi=6.400\rm{GHz}$). In
Fig.~\ref{fig7}(a), we plot Rabi frequency $\Omega_{\rm R}$ as a function
of $\delta=\omega-\Xi_0$. The minimum of the Rabi frequency is not
located at $\delta=0$, but at $\delta \omega/2\pi= 70\rm {MHz}$, which
is very close to the value $\delta\omega/2\pi=66.5 \rm{MHz}$
given in Ref.~\cite{Yoshihara}. One can see that our results are in
good agreement with the numerically calculated data of the flux qubit (NCFQ)
in the whole parameter regime presented in Ref.~\cite{Yoshihara}.
Figure \ref{fig7}(b) displays the
frequency shift as a function of $\Omega_{\rm R0} \equiv
\frac{\Delta}{2}\frac{\rm A}{\Xi_{0}}$ together with the
second-order Bloch-Siegert shift \cite{Shirley,BS-shift}, $\delta
\omega_{\rm BS}^{\rm 2nd}=\frac{1}{4}\frac{\Omega_{\rm  R0}^2}{\Xi_0}$.
It is obvious that $\delta \omega_{\rm BS}^{\rm  2nd}$ overestimates
$\delta \omega$ when $\Omega_{\rm R0}/ 2 \pi \geq
0.8 ~\rm{GHz}$. Our CHRW results are in close agreement with
the results in Ref.~\cite{Yoshihara}. The deviation from the
Bloch-Siegert shift comes from the combined effects of the driving and
the static bias.

We compare in Fig.~\ref{fig7}(c) the generalized Rabi frequency $\Omega_{\rm R}$,
Eq.~(\ref{Rabifreq-CHRW}), as a function of bias $\epsilon$
with the second-order result of Eq.~(\ref{Rabifreq}).
The curves given
by Eq.~(\ref{Rabifreq-CHRW}) and Eq.~(\ref{Rabifreq}) have the same
slope in the parameter regime of Fig.~\ref{fig7}(c).
The generalized Rabi frequency $\Omega_{\rm R}$ versus $\epsilon$ manifests
a linear relation
between them in the resonance case, which can be shown by both
Eq.~(\ref{Rabifreq-CHRW}) and Eq.~(\ref{Rabifreq}).
 We find that the second-order perturbation
results given by Eq.~(\ref{Rabifreq}) agree well with the calculated
result of Fig.~1(c) in Ref. \cite{Yoshihara} obtained by a fit
formula based on the linear approximation.

Figure \ref{fig7}(d) shows
the Rabi frequencies as a function of $\Xi_0$ with a
fixed tunneling strength $\Delta/2\pi=4.869 ~{\rm GHz}$ obtained by three
methods: the CHRW method (the black solid line), the 2nd-order approximation
of the CHRW method (the red dash-dotted line), and the Rabi-RWA method
(the blue dotted line). For a fixed driving frequency
$\omega/2\pi=6.1~{\rm GHz}$, both $\Omega_{\rm R}$ and $\Omega_{\rm R-2nd}$
have a minimum approximately located at $\Xi_0/2\pi=6.4~{\rm GHz}$, near which
the Rabi frequencies are insensitive to the bias
$\epsilon=\sqrt{\Xi_0^2-\Delta^2}$. In contrast, the minimum of
$\Omega_{\rm RR}$ occurs at a position very close to $\Xi_0/2\pi=6.5 ~\rm GHz$, larger
than those of $\Omega_{\rm R}$ and $\Omega_{\rm R-2nd}$. This difference is
attributed to the combined effects of the bias and the driving.

\subsection{ Valid parameter regime of the CHRW method }

In this subsection, we discuss the parameter regime in which the CHRW method is valid in comparison with those of the other approaches. The
Rabi-RWA approach which neglects the CR interactions works well in the weak-driving limit.
The CHRW method covers the the parameter regimes that are good for the Rabi-RWA method which is perturbative in the driving strength \cite{CDT2}.
In the unbiased or small bias case ($\epsilon \ll \Delta$), the CHRW method works very well for $A/\omega \leq 2 $ as shown in Fig. 6 of Ref. \cite{ZG2012}. When $\epsilon \simeq \Delta$, it gives accurate results in the parameter space $A/\omega \leq 1$ regardless of the value of $\omega$. In comparison
with the exact numerical results, the CHRW method works also very well
for the larger bias case $\epsilon \gg \Delta$ (such as
$\epsilon/\Delta \ge 10$) with a fixed driving frequency
$\omega=\Delta$ and the driving strength in the regime of $A/\omega
\le 2$. More interestingly, the CHRW method can give the correct
results when the values of the driving parameters ( $\omega$ and  $A$
) are comparable to those of the energy scales ($\Delta$ and
$\epsilon$). For example, the regime $A\omega/\Delta^2 \leq 1$ in
which the second-order VV method and RWA-RF (see Fig.~\ref{fig5}) fail
is in the valid parameter regime of the CHRW method. Moreover, the
CHRW method works very well also in the parameter regime $A\omega/\Delta^2
\gg 1$ in which the RWA-RF method is valid (see Fig.~\ref{fig5}).

\section{SUMMARY}\label{sec.sum}

We have developed a CHRW method to systematically
investigate the driven dynamics of a TLS under a periodic driving
field and a static bias.
This CHRW method treats
the driving field and the bias on equal footing by a unitary
transformation. The transformed Hamiltonian in the eigenbasis of the
zero-photon Hamiltonian $H'_0$
takes a simple RWA form after we properly choose
the parameters $\xi$ and $\zeta$ in the unitary transformation
by the self-consistent equations (\ref{xi}) and (\ref{zeta}) and
neglect the  $H'_2$ that involves all multi-$\omega$ terms
or multi-photon assisted transitions ($n \omega$, $n=2,3,4...$).
Physically, all the results are dependent on the renormalized energy
splitting $\tilde{\Xi}$ and the renormalized (modified) driving
strength $\tilde{A}$ in
the transformed RWA Hamiltonian. The renormalization of these two
parameters $\tilde{\Xi}$ and $\tilde{A}$ comes from
the combined effect among the tunneling of the TLS, the
driving field and the bias.
The CHRW method allows us to analytically calculate the
driven tunneling dynamics in the renormalized rotating-wave
framework,
so as the generalized bias-modulated Rabi frequency and the frequency shift including
the Bloch-Siegert shift.

We have demonstrated the effects of the driving
field and the bias on the system dynamics and
the generalized Rabi frequency.
We have not only given the small-bias and
weak-driving strength results, such as the Rabi physics, but also shown
the strong driving strength and large bias results, such as the
non-sinusoidally complicated time evolutions.
From the driven tunneling dynamics, one can see that the
characteristics of the oscillations are very
sensitive to $\widetilde A$ and $\widetilde {\Xi}$.
Compared to other analytical methods, the CHRW method
can give the correct and accurate dynamics in good agreement with
the numerically exact results in a broad region of the parameter space
and still preserves the merits of the simple RWA mathematical form in
the final transformed Hamiltonian.
Unlike the
conventional Rabi-RWA method, the CHRW technique is
nonperturbative in driving strength, so it can be applied to study the driven
tunneling physics in a broader parameter regime, especially beyond
the weak driving regime and the small bias and near resonance cases.
In a wide range values of the $\Delta$, $\epsilon$, $A$ and
$\omega$ parameters, we have compared the dynamics obtained by the CHRW
method with that by the numerically exact method. We have found and
confirmed that that the CHRW method works very well even for the
(moderately) strong driving strength region.
The contribution to the dynamics of the multi-$\omega$ terms
or multi-photon assisted transitions ($n \omega$, $n=2,3,4...$) we
have neglected
is not prominent except for the ultra-strong driving strength case.
Thus it gives the accurate driven dynamics in the
parameter regimes of ( $\epsilon/\Delta \ll 1$ or $\gg 1$,
$\Delta/\omega<1$, $A/\omega \leq 2$ ) and ( $\epsilon/\Delta \sim
1 $, $\Delta/\omega \sim 1$, $A/\omega \leq 1$), and in the
neighboring regimes the driven tunneling dynamics though not exact is
in qualitative agreement with that of the numerical method.
By Fourier transform analysis, the discreet frequency values in
$P_{\mathrm{up}}(t)$ of the CHRW results
match very well the ones obtained by the exact numerical method.
More interestingly, the CHRW approach can directly give an
analytical expression for the generalized Rabi frequency
$\Omega_{\rm R}$ and can show explicitly the dependence
of $\Omega_{\rm R}$ on the driving parameters, bias and tunneling.
The results obtained via the CHRW method
might account for the
versatile strongly-driven experiments investigated in
\cite{Fuchs,BS-Exp}.

The CHRW method
provides a simple and direct way for accurately studying the
properties of driven tunneling
systems with a static bias in a wide range of parameters.
The approach is not an upgrade patch
for the conventional RWA but a much improved innovative RWA.
The theoretical results may serve as a tool kit for probing
limitations and possibilities in driven physics and quantum control
with obvious applications in quantum information.
The CHRW method can also be applied to some complicated problems and
dissipative dynamics exposed to strong ac
driving \cite{GH}.
For example, it will prove useful in treating the problems of
multi-chromatically driven tunneling quantum dynamics. The work is
currently under investigation and will appear elsewhere.

\begin{acknowledgments}

Z.L. thanks F. Yoshihara for helpful discussions. The work was
supported by the National Natural Science Foundation of China (Grants
No.~11474200, No.~11581240311, and No.~11374208). Z.L. gratefully
acknowledges support from the National Science Foundation to do
research in ICTP (SMR2705).
H.S.G. acknowledge support from the the Ministry of Science and Technology
of Taiwan under Grant No.~103-2112-M-002-003-MY3, from the National
Taiwan University under Grants No.~NTU-ERP-104R891402, and from the
thematic group program of the National Center for Theoretical Sciences,
Taiwan.

\end{acknowledgments}

\appendix

\section{The occupation probability of the Rabi-RWA \label{sub:Rabi-RWA}}

We discuss
briefly how the Rabi-RWA Hamiltonian is obtained from the Hamiltonian
of Eq.~(\ref{rabi}) and present the occupation probability of the
driven TLS obtained by the Rabi-RWA method.
Writing the transition frequency of the TLS or the flux qubit as
$\Xi_0=\sqrt{\Delta^2+\epsilon^2}$,
thus
we transform the Hamiltonian in the current basis to that in the energy basis by a unitary matrix $U_0 = u_0 \sigma_z - v_0\sigma_x$,
in which
$u_0=\sqrt{ {\frac{1}{2}}-\frac{ \epsilon }{2\Xi_0}}$,
$v_0=\sqrt{ {\frac{1}{2}}+\frac{ \epsilon }{2\Xi_0}}$.
The Hamiltonian in the energy eigenbasis is then written as
$H_{\rm eig}=\frac{\Xi_0}{2}\tilde{\sigma}_z+\frac{A_x}{2}\cos(\omega t)\tilde{\sigma}_x +\frac{A_z}{2}\cos(\omega t) \tilde{\sigma}_z$,
where $\tilde{\sigma}_i$ are the Pauli spin component operators in the
energy basis,
$A_x=(\Delta/\Xi_0) A $ and $A_z= (\epsilon/\Xi_0) A$.
After dropping all the fast oscillatory terms including the last term of
$\frac{A_z}{2}\cos(\omega t) \tilde{\sigma}_z$ in the Hamiltonian $H_{\rm eig}$,
one obtains the Rabi-RWA Hamiltonian
\begin{equation}\label{rwa}
    H_{\mathrm{Rabi-RWA}}=\frac{\Xi_0}{2}\tilde{\sigma}_z + \frac{A_x}{4}
(\tilde{\sigma}_{+}\exp(-i \omega t) + \tilde{\sigma}_{-} \exp(i \omega t))
\end{equation}
where $\tilde{\sigma}_{\pm}=(\tilde{\sigma}_{x}\pm
i\tilde{\sigma}_{y})/2$.  The occupation
probability $P_{\mathrm {up}}(t)$ as an important property is the
probability in the spin-up
eigenstate of the original $\sigma_z$ operator in the Hamiltonian
Eq.(\ref{rabi}) at time $t$ when the system is initialized in the spin-down
eigenstate of the same $\sigma_z$ operator [i.e.,
$P_{\mathrm{up}}(0)=0$]. Since the CR terms
have been neglected in Eq.~(\ref{rwa}), we can
immediately obtain after some algebra the Rabi-RWA
result, $P_{\mathrm {up}}^{\mathrm{Rabi}}(t) = \frac{1+\langle\sigma_z(t)\rangle}{2}$, in which
\begin{eqnarray}\label{pt-rwa}
  \langle\sigma_z(t)\rangle^{\mathrm {Rabi} }&=&-\frac{\epsilon}{\Xi_0}\left\{\frac{\epsilon}{\Xi_0}\left[1-\frac{A_x^2}{2\Omega_{\rm RR}^2} \sin^2\left(\frac{\Omega_{\rm RR} t}{2}\right)\right] + \frac{\Delta}{\Xi_0}\frac{\delta A_x}{\Omega_{\rm RR}^2}\sin^2\left(\frac{\Omega_{\rm RR} t}{2}\right)\right\} \nonumber \\
  &&-\frac{\Delta}{\Xi_0}\left\{ \cos\left(\omega t \right) \left[\frac{\Delta}{\Xi_0}- \left( \frac{\Delta}{\Xi_0} \frac{ 2\delta^2}{\Omega_{\rm RR}^2}-\frac{\epsilon}{\Xi_0} \frac{\delta A_x}{\Omega_{\rm RR}^2}\right)\sin^2\left(\frac{\Omega_{\rm RR} t}{2}\right)\right] \right. \nonumber \\
  &&\left.-\sin(\omega t)\frac{\Delta\delta- \epsilon A_x /2}{\Xi_0 \Omega_{\rm RR}}\sin(\Omega_{\rm RR} t)\right\},
\end{eqnarray}
where $\delta=\Xi_0-\omega$ is the detuning parameter and $\Omega_{\rm
  RR}=\sqrt{\delta^2+(A_x/2)^2}$ denotes the Rabi freqency of the
Rabi-RWA method. To sum up, the Rabi-RWA Hamiltonian Eq. (\ref{rwa}),
which is mathematically simple and analytically solvable, is reduced
from the biased Rabi Hamiltonian Eq.~(\ref{rabi}) after
neglecting all the fast oscillatory terms.
The CHRW method we develop in this paper
is an analytical method that takes into account
the effects of the dropped terms and yet preserves the
mathematical simplicity of the Rabi-RWA method.

\end{document}